\documentclass[aps,prc,showpacs,preprintnumbers,showkeys,twocolumn]{revtex4}
\usepackage{amssymb}
\usepackage{amsmath}
\usepackage{graphicx}
\usepackage{bm}
\usepackage{epsf}
\usepackage{ulem}
\usepackage{color, soul}
\usepackage[dvipsnames]{xcolor}
\newcommand{\tabincell}[2]{
\begin{tabular}{@{}#1@{}}#2\end{tabular}}

\begin{document}

\title{Nuclear symmetry energy and hadron-quark mixed phase in neutron stars}
\author{X. H. Wu}
\affiliation{School of Physics, Nankai University, Tianjin 300071, China}
\author{H. Shen}~\email{shennankai@gmail.com}
\affiliation{School of Physics, Nankai University, Tianjin 300071, China}

\begin{abstract}
We study the hadron-quark mixed phase, which may occur in the interior
of neutron stars. The relativistic mean-field model is employed to describe
the hadronic phase, while the Nambu-Jona-Lasinio model is used
for the quark phase. We examine the effects of nuclear symmetry energy in
the hadronic phase and repulsive vector interaction in the quark phase.
For the treatment of hadron-quark mixed phase, we describe and compare four
methods: (1) energy minimization method, (2) coexisting phases method,
(3) Gibbs construction, and (4) Maxwell construction.
The finite-size effects like surface and Coulomb energies are taken into
account in the energy minimization and coexisting phases methods,
which play a key role in determining the pasta configuration during the
hadron-quark phase transition.
It is found that massive neutron stars may contain hadron-quark pasta phases,
but pure quark matter is unlikely to occur in the interior of neutron stars.

\end{abstract}

\maketitle


\section{Introduction}
\label{sec:1}

Neutron stars are ideal laboratories for the study of dense matter.
In the core of neutron stars, exotic phases like deconfined quarks
may be present~\cite{Heis00,Glen01,Webe05}.
Over the past decades, there have been numerous research works concerning 
the hadron-quark phase transition in neutron stars~\cite{Glen92,Sche99,
Sche00,Latt00,Burg02,Mene03,Shar07,Yang08,Xu10,Chen13,Orsa14}.
In most of the studies, Gibbs construction~\cite{Glen92} and/or Maxwell construction are
commonly used for the description of hadron-quark mixed phases.
In the Maxwell construction, local charge neutrality is imposed,
and furthermore coexisting hadronic and quark phases have equal pressures and
baryon chemical potentials but different electron chemical potentials.
However, in the Gibbs construction, only global charge neutrality is required,
so hadronic and quark phases are allowed to be charged separately and
have continuous chemical potentials.
It is well known that Gibbs and Maxwell constructions correspond respectively
to the two limits of zero and very large surface tension at the hadron-quark interface,
and therefore, the mixed phase with the Gibbs construction has lower energy than
the one with the Maxwell construction. It was reported in
Ref.~\cite{Bhat10} that there are significant differences in the behavior of
compact stars between the Gibbs and Maxwell constructions. It is noticeable that
both Gibbs and Maxwell constructions involve only bulk contributions, where the
finite-size effects like surface and Coulomb energies are neglected.
When surface and Coulomb energies are considered, a hadron-quark mixed
phase with pasta structures is expected to occur~\cite{Heis93,Endo06,Maru07,Yasu14,Spin16,Wu17}.
The geometric configuration in the mixed phase may change from droplet to rod,
slab, tube, and bubble with increasing baryon density~\cite{Endo06,Maru07,Yasu14,Spin16}.
It is interesting to examine how the hadron-quark mixed phase can be affected by
different treatment methods.

To describe hadron-quark pasta phases, we use the Wigner--Seitz approximation,
where the system is divided into periodically repeating charge-neutral cells.
The hadronic and quark phases inside the cell are assumed to be separated
by a sharp interface with a finite surface tension.
It is known that the surface tension plays a key role in determining the structure
of hadron-quark mixed phase~\cite{Heis93,Endo06,Maru07,Yasu14,Spin16,Wu17},
but its value is poorly known so far.
The calculation in the MIT bag model by using the multiple reflection
expansion (MRE) method predicted a value of the surface tension
$\sigma \approx 10$ MeV/fm$^2$~\cite{Berg87}, while recent calculations within
the MRE framework show that the surface tension falls in the range of 2 to 20 MeV/fm$^2$
for baryon densities between 2 to 10 times the nuclear saturation density~\cite{Lugo17,Lugo18}.
A similar calculation in the Nambu-Jona-Lasinio (NJL) model including
color superconductivity yielded $\sigma \approx$145-165 MeV/fm$^2$~\cite{Lugo13}.
The surface tension calculated from a geometric approach fell in
the range $\sigma \approx$7--30 MeV/fm$^2$~\cite{Pint12}.
Due to the uncertainty of $\sigma$, we treat the surface tension
as a free parameter in the present work.
We employ the three-flavor NJL model to describe quark matter,
while the relativistic mean-field (RMF) models are adopted for hadronic matter.
The NJL model has been widely used as an effective theory of QCD for the
description of quark matter~\cite{Mene03,Hats94,Buba05,Lenzi12,Logo13,Bona12},
since it can successfully describe dynamical chiral symmetry breaking
and generation of constituent quark masses.
In this work, we use the NJL model including repulsive vector interactions.
It has been reported in the literature that including repulsive
vector interactions could significantly affect the QCD phase
diagram~\cite{Fuku08,Ueda13,Buba15} and stiffen the equation of state (EOS)
of quark matter~\cite{Yasu14,Spin16,Pagl08,Abuk09,Masu13,Mene14,Hell14,Chu15,Pere16}.

For the description of hadronic matter, we employ two successful RMF
models, namely TM1~\cite{Suga94} and IUFSU~\cite{Fatt10}, which could provide good
descriptions of finite nuclei and acceptable maximum mass of neutron stars.
The TM1 model has been successfully applied in constructing the EOS
for supernova simulations and neutron stars~\cite{Shen02,Shen11}.
The IUFSU model, which was proposed to overcome a smaller neutron-star mass
predicted by the FSU model, has also been used for the study of neutron-star
structure~\cite{Mene11,Bao14a,Bao15}.
Both of these models include nonlinear terms for $\sigma$ and $\omega$ mesons,
while the IUFSU model includes an additional $\omega$-$\rho$ coupling term
that plays a crucial role in determining the density dependence of symmetry
energy and affecting the neutron-star structure~\cite{Fatt10,Horo03,Mene11,Bao14a,Bao15,Prov13}.
It is well known that nuclear symmetry energy $E_{\textrm{sym}}$ and its
slope $L$ play an important role in understanding many phenomena in nuclear
physics and astrophysics~\cite{LiBA08,Horo01,Oyam07,Duco10,Zhan13,Bald16,Oert17}.
Especially, neutron-star radii and crust structures are closely related to
the symmetry energy slope $L$~\cite{Mene11,Bao14a,Bao15,Prov13}.
Recently, many efforts have been devoted to constraining the values
of $E_{\rm sym}$ and $L$ at saturation density based
on astrophysical observations and terrestrial nuclear
experiments~\cite{Tews17,Hebe13,Latt14,Hage15,Dani14,Roca15,Birk17,Dani17}.
In Refs.~\cite{Oert17,Tews17}, a sufficient number of constraints on the symmetry
energy parameters have been summarized and the most probable values
for the symmetry energy and its slope at saturation density
were found to be $E_{\rm sym}=31.7\pm 3.2$ MeV and $L=58.7\pm 28.1$ MeV, respectively,
with a much larger error for $L$ than that for $E_{\rm sym}$.
Although the TM1 and IUFSU models give similar binding energies of finite
nuclei, their symmetry energy slopes are very different from each
other ($L=47.2$ MeV in IUFSU and $L=110.8$ MeV in TM1).
In order to examine the $L$ dependence of hadron-quark pasta phases, we
employ two sets of generated models based on the TM1 and IUFSU parametrizations
as described in Ref.~\cite{Bao14b}. The models in each set were obtained by
simultaneously adjusting $g_{\rho}$ and ${\Lambda}_{\rm{v}}$ so as to achieve
a given $L$ at saturation density $n_0$ while keeping $E_{\rm{sym}}$ fixed
at the baryon density $n_b=0.11\, \rm{fm}^{-3}$. The choice of fixing symmetry
energy at $n_b=0.11\, \rm{fm}^{-3}$ aims to produce similar binding energies
of finite nuclei within one set of generated models, which should
be consistent with experimental data.
It is noticeable that all models in each set have the same
isoscalar saturation properties and fixed symmetry energy $E_{\rm{sym}}$ at
$n_b=0.11\, \rm{fm}^{-3}$ but have different symmetry energy slope $L$.
By using a set of models with different values of $L$, it is possible to
study the effects of nuclear symmetry energy on the hadron-quark phase
transition and pasta structures in neutron stars.

The main purpose of this article is to investigate the properties of hadron-quark
pasta phases, which could be affected by nuclear symmetry energy and other
parameters such as the surface tension $\sigma$ and vector coupling $G_V$ in
the NJL model. To describe the hadron-quark pasta phase, we use the energy
minimization (EM) method, where the equilibrium state at a given baryon density
is determined by minimizing the total energy density. The EM method, which is
referred to as the compressible liquid-drop (CLD) model, has been widely used
in the study of nuclear liquid-gas phase transition at subnuclear
densities~\cite{Bao14b,Baym71,Latt85,Latt91}. In the EM method,
the equilibrium conditions for coexisting phases are
derived by minimization of the total energy including surface and
Coulomb contributions, which are different from the Gibbs conditions
without finite-size effects. Furthermore, a simple coexisting
phases (CP) method~\cite{Bao14a,Mene08} is also used and compared for the
description of hadron-quark pasta phases. In the CP method, two coexisting phases
satisfy the Gibbs conditions for phase equilibrium, which require equal pressures
and chemical potentials for two
phases~\cite{Glen92,Sche99,Sche00,Latt00,Burg02,Mene03,Shar07,Yang08,Xu10}.
After the equilibrium state is obtained by applying the Gibbs conditions,
the surface and Coulomb energies are perturbatively taken into account in the CP method.
Since the equilibrium conditions in the EM method are derived
by minimization of the total energy including surface and Coulomb
contributions, the finite-size effects are treated relatively well
in the EM method. By comparing results from different treatments, we
can examine how the pasta structures could be influenced by the method used
in the calculations.

This article is organized as follows.
In Sec.~\ref{sec:2}, we briefly describe the RMF models for hadronic matter and discuss
the choice of parameters.
In Sec.~\ref{sec:3}, the NJL model used for quark matter is briefly introduced.
In Sec.~\ref{sec:4}, we describe and compare the four methods used for the study
of hadron-quark mixed phase, namely (1) EM method, (2) CP method, (3) Gibbs construction, and
(4) Maxwell construction.
In Sec.~\ref{sec:5}, we show numerical results and discuss the properties
of hadron-quark mixed phase in neutron stars.
The effects of nuclear symmetry energy and model dependence are also examined.
Section~\ref{sec:6} is devoted to the conclusions.

\section{Hadronic matter phase}
\label{sec:2}

The hadronic matter is described by the RMF model, where nucleons interact
via the exchange of isoscalar-scalar meson $\sigma$, isoscalar-vector meson $\omega$,
and isovector-vector meson $\rho$.
For hadronic matter consisting of nucleons ($p$ and $n$) and
leptons ($e$ and $\mu$), the Lagrangian density is written as
\begin{eqnarray}
\label{eq:LRMF}
\mathcal{L}_{\rm{RMF}} & = & \sum_{i=p,n}\bar{\psi}_i
\left\{i\gamma_{\mu}\partial^{\mu}-\left(M+g_{\sigma}\sigma\right) \right.\notag \\
&& \left. -\gamma_{\mu} \left[g_{\omega}\omega^{\mu} +\frac{g_{\rho}}{2}\tau_a\rho^{a\mu}
\right]\right\}\psi_i  \notag \\
&& +\frac{1}{2}\partial_{\mu}\sigma\partial^{\mu}\sigma -\frac{1}{2}%
m^2_{\sigma}\sigma^2-\frac{1}{3}g_{2}\sigma^{3} -\frac{1}{4}g_{3}\sigma^{4}
\notag \\
&& -\frac{1}{4}W_{\mu\nu}W^{\mu\nu} +\frac{1}{2}m^2_{\omega}\omega_{\mu}%
\omega^{\mu} +\frac{1}{4}c_{3}\left(\omega_{\mu}\omega^{\mu}\right)^2  \notag
\\
&& -\frac{1}{4}R^a_{\mu\nu}R^{a\mu\nu} +\frac{1}{2}m^2_{\rho}\rho^a_{\mu}%
\rho^{a\mu} \notag \\
&& +\Lambda_{\rm{v}} \left(g_{\omega}^2
\omega_{\mu}\omega^{\mu}\right)
\left(g_{\rho}^2\rho^a_{\mu}\rho^{a\mu}\right) \notag\\
&& +\sum_{l=e,\mu}\bar{\psi}_{l}
  \left( i\gamma_{\mu }\partial^{\mu }-m_{l}\right)\psi_l,
\end{eqnarray}
where $W^{\mu\nu}$ and $R^{a\mu\nu}$ are the antisymmetric field
tensors corresponding to $\omega^{\mu}$ and $\rho^{a\mu}$, respectively.
In the RMF approach, the meson fields are treated as classical fields and the field
operators are replaced by their expectation values. For a static system,
the nonvanishing expectation values are $\sigma =\left\langle \sigma \right\rangle$,
$\omega =\left\langle\omega^{0}\right\rangle$, and $\rho =\left\langle \rho^{30} \right\rangle$.

In uniform hadronic matter, the equations of motion for meson mean fields have the following form:
\begin{eqnarray}
m_{\sigma }^{2}\sigma +g_{2}\sigma ^{2}+g_{3}\sigma^{3}
&=&-g_{\sigma }\left( n_{p}^{s}+n_{n}^{s}\right) ,
\label{eq:eqms} \\
m_{\omega }^{2}\omega +c_{3}\omega^{3}
+2\Lambda_{\rm{v}}g^2_{\omega}g^2_{\rho}{\rho}^2 \omega
&=&g_{\omega}\left( n_{p}+n_{n}\right) ,
\label{eq:eqmw} \\
m_{\rho }^{2}{\rho}
+2\Lambda_{\rm{v}}g^2_{\omega}g^2_{\rho}{\omega}^2{\rho}
&=&\frac{g_{\rho }}{2}\left(n_{p}-n_{n}\right) ,
\label{eq:eqmr}
\end{eqnarray}%
where $n_i^s$ and $n_i$ denote the scalar and number densities
of species $i$, respectively. The energy density of hadronic matter
is given by
\begin{eqnarray}
\varepsilon_{\rm{HP}} &=&\sum_{i=p,n}\frac{1}{\pi^2}
     \int_{0}^{k^{i}_{F}}{\sqrt{k^2+{M^{\ast}}^2}}\ k^2dk   \nonumber \\
&& + \frac{1}{2}m^2_{\sigma}{\sigma}^2+\frac{1}{3}{g_2}{\sigma}^3
     +\frac{1}{4}{g_3}{\sigma}^4  + \frac{1}{2}m^2_{\omega}{\omega}^2 \nonumber  \\
&&   + \frac{3}{4}{c_3}{\omega}^4
     + \frac{1}{2}m^2_{\rho}{\rho}^2
     + 3{\Lambda}_{\textrm{v}}\left(g^2_{\omega}{\omega}^2\right)
     \left(g^2_{\rho}{\rho}^2\right) \nonumber  \\
&& + \sum_{l=e,\mu}\frac{1}{\pi^{2}}\int_{0}^{k_{F}^{l}}
     \sqrt{k^{2}+m_{l}^{2}}\ k^{2}dk,
     \label{eq:ehp}
\end{eqnarray}
and the pressure is written as
\begin{eqnarray}
P_{\rm{HP}} &=& \sum_{i=p,n}\frac{1}{3\pi^2}\int_{0}^{k^{i}_{F}}
      \frac{k^4dk}{\sqrt{k^2+{M^{\ast}}^2}}   \nonumber  \\
&& - \frac{1}{2}m^2_{\sigma}{\sigma}^2-\frac{1}{3}{g_2}{\sigma}^3
     -\frac{1}{4}{g_3}{\sigma}^4  + \frac{1}{2}m^2_{\omega}{\omega}^2 \nonumber \\
&&    +\frac{1}{4}{c_3}{\omega}^4
      + \frac{1}{2}m^2_{\rho}{\rho}^2 +
      \Lambda_{\textrm{v}}\left(g^2_{\omega}{\omega}^2\right)
      \left(g^2_{\rho}{\rho}^2\right) \nonumber \\
&& + \sum_{l=e,\mu}\frac{1}{3\pi^{2}}\int_{0}^{k_{F}^{l}}
     \frac{k^{4} dk}{\sqrt{k^{2}+m_{l}^{2}}},
\label{eq:php}
\end{eqnarray}
where $M^{\ast}=M+g_{\sigma}{\sigma}$ denotes the effective nucleon mass.
For hadronic matter in $\beta$ equilibrium, the chemical potentials
satisfy the relations $\mu_{p}=\mu_{n}-\mu_{e}$ and $\mu_{\mu}=\mu_{e}$.
At zero temperature, the chemical potentials are given by
\begin{eqnarray}
\mu_i &=& \sqrt{{k_{F}^{i}}^{2}+{M^{\ast}}^2}+g_{\omega}\omega
+g_{\rho}\tau_{3}^{i}\rho, \hspace{0.5cm}  i=p,n,\\
\mu_{l} &=& \sqrt{{k_{F}^{l}}^{2}+m_{l}^{2}}, \hspace{3.0cm}  l=e,\mu.
\end{eqnarray}

In order to investigate the impact of nuclear symmetry energy on the hadron-quark
phase transition, we adopt two successful RMF models, TM1~\cite{Suga94},
and IUFSU~\cite{Fatt10}, to describe nuclear interactions.
For completeness, we present the parameter sets and saturation properties
of these two models in Tables~\ref{tab:para} and~\ref{tab:satu}, respectively.
It is well known that nuclear symmetry energy $E_{\textrm{sym}}$ and its
slope $L$ play a crucial role in determining the properties of neutron stars.
To examine the $L$ dependence of hadron-quark pasta phases, we employ two sets
of generated models based on the TM1 and IUFSU parametrizations as
described in Ref.~\cite{Bao14b}. We note that all models in each set have the same
isoscalar saturation properties and fixed symmetry energy $E_{\rm{sym}}$ at a
density of $0.11\, \rm{fm}^{-3}$ but have different symmetry energy slope $L$.
It has been reported in Ref.~\cite{Bao14b} that the choice of fixing symmetry energy
at $n_b=0.11\, \rm{fm}^{-3}$ could produce very similar binding energies of finite nuclei
within one set of generated models.  The generated models were obtained
by simultaneously adjusting $g_{\rho}$ and ${\Lambda}_{\rm{v}}$
so as to achieve a given $L$ at saturation density $n_0$ while
keeping $E_{\rm{sym}}$ fixed at $n_b=0.11\, \rm{fm}^{-3}$.
The parameters, $g_{\rho}$ and ${\Lambda}_{\rm{v}}$, generated from
the TM1 and IUFSU models for different $L$ are given in
Tables~\ref{tab:tm1-L} and~\ref{tab:iufsu-L} for completeness.
For the TM1 model, we consider that $L$ varies from $50$ 
to $110.8$ MeV (original TM1 value).
For the IUFSU model, the range of $L$ is from $47.2$ (original IUFSU value)
to $110$ MeV.
As one can see in Tables~\ref{tab:tm1-L} and~\ref{tab:iufsu-L}, there is
a positive correlation between the slope parameter $L$ and the symmetry energy
at saturation density $E_{\rm{sym}}(n_0)$. In the case of TM1,
we obtain $E_{\rm{sym}}(n_0)=32.39$ MeV for $L=50$ MeV,
while $E_{\rm{sym}}(n_0)=36.89$ MeV for $L=110.8$ MeV.
\begin{table*}[htbp]
\caption{Parameter sets TM1 and IUFSU for the RMF Lagrangian. All masses are in MeV.}
\label{tab:para}
\setlength{\tabcolsep}{2.4mm}{
\begin{tabular}{lccccccccccc}
\hline\hline
Model   &$M$  &$m_{\sigma}$  &$m_\omega$  &$m_\rho$  &$g_\sigma$  &$g_\omega$
        &$g_\rho$ &$g_{2}$ (fm$^{-1}$) &$g_{3}$ &$c_{3}$ &$\Lambda_{\textrm{v}}$ \\
\hline
TM1     &938.0  &511.198  &783.0  &770.0  &10.0289  &12.6139  &9.2644
        &$-$7.2325   &0.6183   &71.3075   &0.000  \\
IUFSU   &939.0  &491.500  &782.5  &763.0  &9.9713   &13.0321  &13.5900
        &$-$8.4929   &0.4877   &144.2195  &0.046 \\
\hline\hline
\end{tabular}}
\end{table*}
\begin{table}[htbp]
\caption{Saturation properties of symmetric nuclear matter for the TM1 and IUFSU models.
The quantities $E_0$, $K$, $E_{\text{sym}}$, and $L$ are, respectively,
the energy per nucleon, incompressibility coefficient, symmetry
energy, and symmetry energy slope at saturation density $n_0$.}
\label{tab:satu}
{\footnotesize
\begin{center}
\begin{tabular*}{1\linewidth}{lccccc}
\hline\hline
 Model & $n_0$ (fm$^{-3}$) & $E_0$ (MeV) & $K$ (MeV) & $E_{\text{sym}}$ (MeV) & $L$ (MeV) \\
\hline
TM1   & 0.145 & $-$16.3 & 281 & 36.9 & 110.8 \\
IUFSU & 0.155 & $-$16.4 & 231 & 31.3 & 47.2  \\
\hline\hline
\end{tabular*}
\end{center}}
\end{table}
\begin{table*}[htbp]
\caption{Parameters $g_\rho$ and $\Lambda_{\text{v}}$, generated from the TM1 model
for different slope $L$ at saturation density $n_0$ with fixed symmetry energy
$E_{\text{sym}}=28.05$ MeV at a density of $0.11\, \rm{fm}^{-3}$.
The last line shows the symmetry energy at saturation density $n_0$.
The original TM1 model has $L=110.8$ MeV.}
\label{tab:tm1-L}
\begin{center}
\setlength{\tabcolsep}{5.2mm}{
\begin{tabular}{lccccccc}
\hline\hline
$L$ (MeV) & 50.0    & 60.0    & 70.0    & 80.0    & 90.0   & 100.0  & 110.8  \\
\hline
$g_\rho$  & 12.2413 & 11.2610 & 10.6142 & 10.1484 & 9.7933 & 9.5114 & 9.2644 \\
$\Lambda_{\text{v}}$ & 0.0327  & 0.0248  & 0.0182  & 0.0128 & 0.0080 & 0.0039 & 0.0000 \\
$E_{\rm{sym}}(n_0)$ (MeV) & 32.39 & 33.29 & 34.11 & 34.86 & 35.56 & 36.22 & 36.89 \\
\hline\hline
\end{tabular}}
\end{center}
\end{table*}
\begin{table*}[htbp]
\caption{Parameters $g_\rho$ and $\Lambda_{\text{v}}$, generated from the IUFSU model
for different slope $L$ at saturation density $n_0$ with fixed symmetry energy
$E_{\text{sym}}=26.78$ MeV at a density of $0.11\, \rm{fm}^{-3}$.
The last line shows the symmetry energy at saturation density $n_0$.
The original IUFSU model has $L=47.2$ MeV.}
\label{tab:iufsu-L}
\begin{center}
\setlength{\tabcolsep}{4.3mm}{
\begin{tabular}{lcccccccc}
\hline\hline
$L$ (MeV) & 47.2    & 50.0    & 60.0    & 70.0    & 80.0   & 90.0   & 100.0  & 110.0  \\
\hline
$g_\rho$  & 13.5900 & 12.8202 & 11.1893 & 10.3150 & 9.7537 & 9.3559 & 9.0558 & 8.8192 \\
$\Lambda_{\text{v}}$& 0.0460  & 0.0420  & 0.0305  & 0.0220 & 0.0153 & 0.0098 & 0.0051 & 0.0011 \\
$E_{\rm{sym}}(n_0)$ (MeV) & 31.30 & 31.68 & 32.89 & 33.94 & 34.88 & 35.74 & 36.53 & 37.27 \\
\hline\hline
\end{tabular}}
\end{center}
\end{table*}

\section{Quark matter phase}
\label{sec:3}

To describe quark matter, we adopt the NJL model with three flavors.
The Lagrangian density of the NJL model is give by
\begin{eqnarray}
\label{eq:Lnjl}
\mathcal{L}_{\rm{NJL}} &=&\bar{q}\left( i\gamma _{\mu }\partial ^{\mu
}-m^{0}\right) q \nonumber \\
&&+{G_S}\sum\limits_{a = 0}^8 {\left[ {{{\left( {\bar q{\lambda _a}q} \right)}^2}
+ {{\left( {\bar q i{\gamma _5}{\lambda _a}q} \right)}^2}} \right]}  \nonumber \\
&&-K\left\{ \det \left[ \bar{q}\left( 1+\gamma _{5}\right) q\right] +\det %
\left[ \bar{q}\left( 1-\gamma _{5}\right) q\right] \right\} \nonumber \\
&&- {G_V}\sum\limits_{a = 0}^8 {\left[ {{{\left( {\bar q{\gamma ^\mu }{\lambda _a}q} \right)}^2}
+ {{\left( {\bar q{\gamma ^\mu }{\gamma _5}{\lambda _a}q} \right)}^2}} \right]},
\end{eqnarray}%
where $q$ denotes the quark field with three flavors ($N_f=3$) and three colors ($N_c=3$).
The current quark mass matrix is given by
$m^{0}=\text{diag} \left(m_{u}^{0},m_{d}^{0},m_{s}^{0}\right)$.
We take into account chirally symmetric four-quark interaction with coupling ${G_S}$,
Kobayashi--Maskawa--'t Hooft (KMT) six-quark interaction with coupling ${K}$, and vector
interaction with coupling ${G_V}$. It has been shown in
Refs.~\cite{Yasu14,Spin16,Pagl08,Abuk09,Masu13,Mene14,Hell14,Chu15,Pere16}
that vector interactions in the NJL model play an important role in describing
massive stars. In the present work, we use the parameters given in Ref.~\cite{Rehb96},
$m_{u}^{0}=m_{d}^{0}=5.5\ \text{MeV}$, $m_{s}^{0}=140.7\ \text{MeV}$,
$\Lambda =602.3\ \text{MeV}$, ${G_S}\Lambda^{2}=1.835$,
and $K\Lambda ^{5}=12.36$. The vector coupling ${G_V}$ is treated as a free
parameter following Refs.~\cite{Yasu14,Spin16,Pere16}, since there is still no constraint on
${G_V}$ at finite density. In Ref.~\cite{Yasu14}, two values were used for the
ratio ${G_V}/{G_S}=0.1$ and $0.2$. Several values of ${G_V}/{G_S}$
between 0 to 0.75 were adopted in Ref.~\cite{Pere16}. In the present work
we find that, using the TM1 model for the hadronic phase, a pure quark phase does not appear
at densities below 20 times
the nuclear saturation density for ${G_V}/{G_S} > 0.45$, which implies that
such a large ${G_V}$ is not suggested.
Therefore, we use ${G_V}/{G_S}=0$ and $0.4$ to investigate the effects of vector
couplings in the present calculations, where the TM1 model is employed to describe the hadronic phase.
Since ${G_V}$ can only stiffen the EOS of quark matter, the effects of vector
couplings on the hadron-quark phase transition using the IUFSU model would be qualitatively
similar to the case of using the TM1 model. Furthermore, the onset density of the mixed phase obtained
in the IUFSU model is significantly higher than that of the TM1 model (see Fig.~\ref{fig:3enb} below),
and therefore we use only ${G_V}/{G_S}=0$ in the calculations with the IUFSU model.

At the mean-field level, the quarks get constituent quark masses by spontaneous
chiral symmetry breaking. The constituent quark mass
in vacuum $m_{i}$ is much larger than the current quark mass $m_{i}^{0}$.
The constituent quark masses $m_{i}^{\ast }$ in quark matter can be determined
by the gap equations
\begin{equation}
\label{eq:gap}
m_{i}^{\ast }=m_{i}^{0}-4{G_S}\langle \bar{q}_{i}q_{i}\rangle +2K\langle \bar{q}%
_{j}q_{j}\rangle \langle \bar{q}_{k}q_{k}\rangle,
\end{equation}%
with ($i,j,k$) being any permutation of ($u,d,s$).
The energy density of quark matter is given by
\begin{eqnarray}
\label{eq:eNJL}
\varepsilon_{\rm{NJL}} &=&\sum\limits_{i = u,d,s}
 {\left[ { - \frac{3}{{{\pi ^2}}}\int_{k_F^i}^\Lambda
  {\sqrt {{k^2} + m_i^{ * 2}} } \;{k^2}dk} \right]} \notag\\
 & &  + 2{G_S}\left( {C_u^2 + C_d^2 + C_s^2} \right) - 4K{C_u}{C_d}{C_s} \notag\\
 & &  + 2{G_V}\left( {n_u^2 + n_d^2 + n_s^2} \right)   - {\varepsilon _0},
\end{eqnarray}%
where $C_{i}=\left\langle \bar{q}_{i}q_{i}\right\rangle $ denotes the quark condensate
of flavor $i$.  The constant $\varepsilon_{0}$ is introduced to set $\varepsilon_{\rm{NJL}}=0$ in
the physical vacuum. In Refs.~\cite{Lenzi12,Logo13,Bona12}, an effective bag constant
$B^*$ was introduced since there remains uncertainty in the low-density
normalization of pressure in the NJL model.
In the present work, our choice of $\varepsilon_{0}$ corresponds to
a vanishing pressure in the vacuum.

The chemical potentials of quarks and leptons in quark matter satisfy
the $\beta$ equilibrium condition,
$\mu_{s}=\mu_{d}=\mu_{u}+\mu_{e}$ and $\mu_{\mu}=\mu_{e}$, where
the chemical potential of quark flavor $i$ is given by
\begin{eqnarray}
\mu_i =\sqrt{{k_{F}^{i}}^{2}+{m_{i}^{\ast}}^{2}}+ 4 G_V \, n_i.
\end{eqnarray}
The total energy density and pressure in quark matter
are written as
\begin{eqnarray}
\label{eq:e2}
\varepsilon_{\rm{QP}} &=& \varepsilon_{\rm{NJL}}
  +\sum_{l=e,\mu }\frac{1}{\pi^{2}}
\int_{0}^{k_{F}^{l}}\sqrt{k^{2}+m_{l}^{2}}\ k^{2}dk,
\\
P_{\rm{QP}} &=&\sum_{i=u,d,s,e,\mu }n_{i}\mu_{i}-\varepsilon_{\rm{QP}}.
\label{eq:p2}
\end{eqnarray}

\section{Hadron-quark mixed phase}
\label{sec:4}

In this section, we briefly introduce and compare several methods for
the description of hadron-quark mixed phase, namely (1) energy minimization (EM) method,
(2) coexisting phases (CP) method, (3) Gibbs construction, and (4) Maxwell construction.
We emphasize that the main difference among these methods is the treatment of surface and
Coulomb contributions. Generally, the hadron-quark mixed phase can be described by
the Wigner--Seitz approximation, where the system is divided into periodically repeating
charge-neutral cells. The coexisting hadronic and quark phases inside the cell are separated by
a sharp interface where a surface tension often exists.
The possible geometric structure of the mixed phase may change
from droplet to rod, slab, tube, and bubble with increasing baryon density.
In the EM method, the equilibrium conditions between coexisting hadronic and quark phases
are determined by minimization of the total energy including surface and Coulomb
contributions, so the finite-size effects due to surface and Coulomb contributions
are treated relatively well compared with other methods.
In the CP method, the surface and Coulomb energies are perturbatively included,
while the Gibbs equilibrium conditions are used for the two coexisting phases.
We note that both the Gibbs and Maxwell constructions do not include the finite-size effects.
In the Gibbs construction, the surface tension at the hadron-quark interface is assumed to
be negligible, hence the surface and Coulomb energies are not taken into account
and only global charge neutrality is required. On the other hand, the surface tension
in the Maxwell construction is assumed to be extremely large,
so that local charge neutrality has to be maintained.
The surface tension plays a key role in determining the structure of hadron-quark
mixed phase, but its value is poorly known so far.
In the present work, we treat the surface tension $\sigma$ as a free parameter.

In the following subsections, we describe how to determine the equilibrium state
of hadron-quark mixed phase at a given baryon density within different methods.

\subsection{Energy minimization method}
\label{sec:4-1}

The Wigner--Seitz approximation is used to describe the hadron-quark mixed phase,
where two coexisting phases inside a charge-neutral cell are separated
by a sharp interface with a finite surface tension. The leptons (electrons and muons)
are assumed to be uniformly distributed throughout the cell.
The total energy density of the mixed phase is given by
\begin{eqnarray}
\label{eq:fws}
\varepsilon_{\rm{MP}} &=& u \varepsilon_{\rm{QP}}
  + (1 - u)\varepsilon_{\rm{HP}}
  + \varepsilon_{\rm{surf}} + \varepsilon_{\rm{Coul}} ,
\end{eqnarray}%
where $u=V_{\rm{QP}}/(V_{\rm{QP}}+V_{\rm{HP}})$ is the volume fraction
of the quark phase. The energy densities, $\varepsilon_{\rm{HP}}$ and $\varepsilon_{\rm{QP}}$,
are given by Eqs.~(\ref{eq:ehp}) and (\ref{eq:e2}), respectively.
The surface and Coulomb energy densities are expressed as
\begin{eqnarray}
{\varepsilon}_{\rm{surf}}
&=& \frac{D \sigma u_{\rm{in}}}{r_D},
\label{eq:esurf} \\
{\varepsilon}_{\textrm{Coul}}
&=& \frac{e^2}{2}\left(\delta n_c\right)^{2}r_D^{2} u_{\rm{in}}\Phi\left(u_{\rm{in}}\right),
\label{eq:ecoul}
\end{eqnarray}%
with%
\begin{eqnarray}
\label{eq:Du}
\Phi\left(u_{\rm{in}}\right)=\left\{
\begin{array}{ll}
\frac{1}{D+2}\left(\frac{2-Du_{\rm{in}}^{1-2/D}}{D-2}+u_{\rm{in}}\right),  & D=1,3, \\
\frac{u_{\rm{in}}-1-\ln{u_{\rm{in}}}}{D+2},  & D=2. \\
\end{array} \right.
\end{eqnarray}%
Here, $\sigma$ denotes the surface tension at the hadron-quark interface,
while $D=1,2,3$ is the geometric dimension of the cell with $r_D$ being the size
of the inner part. $u_{\rm{in}}$ represents the volume fraction of the inner part,
i.e., $u_{\rm{in}}=u$ for droplet, rod, and slab configurations,
and $u_{\rm{in}}=1-u$ for tube and bubble configurations.
$e=\sqrt{4\pi/137}$ is the electromagnetic coupling constant.
$\delta n_c=n_c^{\rm{HP}}-n_c^{\rm{QP}}$ is the charge-density
difference between hadronic and quark phases.
In Eq.~(\ref{eq:fws}), the first two terms represent the bulk contributions,
while the last two terms come from the finite-size effects that depend on
the size $r_D$. At a given baryon density, $r_D$ can be determined by minimizing
${\varepsilon}_{\rm{surf}}+{\varepsilon}_{\rm{Coul}}$, which leads to the relation
${\varepsilon}_{\rm{surf}}=2{\varepsilon}_{\rm{Coul}}$.
The size of the inner phase and that of the Wigner--Seitz cell are respectively
given by
\begin{eqnarray}
\label{eq:rd}
r_D &=& \left[\frac{\sigma{D}}{e^2
\left(\delta n_c\right)^{2}\Phi}\right]^{1/3}, \\
\label{eq:rc}
r_C &=&   u^{-1/D} r_D.
\end{eqnarray}

In the EM method, the equilibrium conditions for coexisting hadronic and quark
phases in the Wigner--Seitz cell are derived by minimization of the total energy
density~(\ref{eq:fws}). At a given baryon density $n_b$, the energy density of the mixed
phase $\varepsilon_{\rm{MP}}$ is considered as a function of eight variables:
$n_{p}$, $n_{n}$, $n_{u}$, $n_{d}$, $n_{s}$, $n_{e}$, $n_{\mu}$, and $u$.
The minimization should be performed under the constraints of globe charge neutrality
and baryon number conservation, which are expressed as,
\begin{eqnarray}
\label{eq:nc}
0 &=& n_e + n_\mu -\frac{u}{3}\left( 2n_u - n_d - n_s \right)
  - (1 - u)  n_p   , \\
n_b &=& \frac{u}{3}\left( n_u + n_d + n_s \right)
    + (1 - u) \left( n_p + n_n \right).
\label{eq:nb}
\end{eqnarray}
We introduce the Lagrange multipliers, $\mu_e$ and $\mu_n$,
for the constraints, and then construct a function as
{\small \begin{eqnarray}
w &=&\varepsilon_{\rm{MP}}
  -\mu_e \left[ n_e + n_\mu - \frac{u}{3}(2n_u - n_d - n_s)
    - (1 - u)  n_p  \right] \notag \\
  & & -\mu_n \left[ \frac{u}{3}\left( n_u + n_d + n_s \right) + (1 - u) \left( n_p + n_n \right)\right].
\end{eqnarray}}
By minimizing $w$ with respect to the particle densities, we obtain
the following equilibrium conditions for chemical potentials:
\begin{eqnarray}
\mu_u - \frac{ 4\varepsilon_{\rm{Coul}} }{3u \, \delta n_c}
  &=& \frac{1}{3}\mu_n - \frac{2}{3}\mu_e, \label{eq:CU1}\\
\mu_d + \frac{ 2\varepsilon_{\rm{Coul}} }{3u \, \delta n_c}
  &=& \frac{1}{3}\mu_n + \frac{1}{3}\mu_e, \label{eq:CD1}\\
\mu_s + \frac{ 2\varepsilon_{\rm{Coul}} }{3u \, \delta n_c}
  &=& \frac{1}{3}\mu_n + \frac{1}{3}\mu_e, \label{eq:CS1}\\
\mu_p +\frac{ 2\varepsilon_{\rm{Coul}} }{(1-u) \, \delta n_c}
  &=& \mu_n - \mu_e, \label{eq:CN1}\\
\mu_\mu &=& \mu_e. \label{eq:CE1}
\end{eqnarray}
The equilibrium condition for the pressure at the interface is achieved by minimizing
$w$ with respect to the volume fraction $u$, which can be written as
\begin{eqnarray}
P_{\rm{HP}} &=& P_{\rm{QP}} -\frac{2\varepsilon_{\rm{Coul}}}{\delta n_c}
  \left[ \frac{1}{3u}\left( 2n_u - n_d - n_s \right)+\frac{1}{1-u}n_p\right] \notag \\
& &  \mp \frac{\varepsilon_{\rm{Coul}} }{u_{\rm{in}}}
  \left(3+u_{\rm{in}}\frac{{\Phi}^{^{\prime }}}{\Phi}\right),
\label{eq:CP1}
\end{eqnarray}
where the sign of the last term is $-$ for droplet, rod, and slab configurations,
while it is $+$ for tube and bubble configurations.
The equilibrium equations~(\ref{eq:CU1})--(\ref{eq:CP1}) are clearly
different from the Gibbs equilibrium conditions, which is due to the inclusion
of surface and Coulomb energies in the minimization procedure.
We define the pressure of the mixed phase by the thermodynamic relation,
$P_{\rm{MP}} = n_b^{2}\frac{\partial \left(\varepsilon_{\rm{MP}}/n_b\right) }
{\partial n_b}$, which is somewhat different from $P_{\rm{HP}}$ and $P_{\rm{QP}}$.
This is similar to the case of nuclear liquid-gas phase transition
at subnuclear densities~\cite{Bao14b,Baym71,Latt85,Latt91}.

By solving the above equilibrium equations at a given baryon density $n_b$,
we calculate and compare the energy density of the mixed phase with different
pasta configurations, and then determine the most stable shape that has the
lowest energy density. All thermodynamic quantities of the mixed phase
can be computed after the equilibrium state is achieved.

\subsection{Coexisting phases method}
\label{sec:4-2}

In the CP method, the Gibbs equilibrium conditions are used for coexisting hadronic
and quark phases. Meanwhile, the surface and Coulomb energies are included perturbatively.
This means that the Gibbs equilibrium conditions are derived without the inclusion of
surface and Coulomb contributions, but they are taken into account in the
total energy density of the mixed phase given by Eq.~(\ref{eq:fws}).
In fact, we can derive the Gibbs equilibrium conditions by minimizing
the total energy density without surface and Coulomb terms.
By setting ${\varepsilon}_{\rm{surf}}={\varepsilon}_{\rm{Coul}}=0$ in Eq.~(\ref{eq:fws}),
we minimize the energy density following the procedure described in the EM method.
The resulting equilibrium conditions are given by,
\begin{eqnarray}
P_{\rm{HP}} &=& P_{\rm{QP}}, \label{eq:CP2} \\
\mu_u+\mu_e &=& \mu_d = \mu_s = \frac{1}{3}\mu_n + \frac{1}{3}\mu_e, \label{eq:CU2}\\
\mu_p &=& \mu_n - \mu_e, \label{eq:CN2}\\
\mu_\mu &=& \mu_e, \label{eq:CE2}
\end{eqnarray}
which are equivalent to the Gibbs conditions for phase equilibrium in Ref.~\cite{Yang08}.
After the equilibrium state is achieved by solving Eqs.~(\ref{eq:CP2})--(\ref{eq:CE2})
at a given baryon density $n_b$, the energy density of the mixed phase can be calculated
from Eq.~(\ref{eq:fws}), where the surface and Coulomb energies are taken into account.
Compared with the EM method, the CP method can be viewed as a perturbative
approximation, in which the surface and Coulomb contributions are regarded as perturbations
and their influences on the equilibrium conditions are neglected.
Therefore, the EM method is more complete than the CP method, because the surface and Coulomb
contributions are included not only in the total energy density of the mixed phase but also in
the equilibrium conditions for coexisting phases in the EM method.
We emphasize that the shape and size of the mixed phase are
determined by competition between surface and Coulomb energies,
which are unrelated to the Gibbs conditions.
In the CP method, the pressure of the mixed phase satisfies the
relation $P_{\rm{MP}} = P_{\rm{HP}} = P_{\rm{QP}}$.
This is because the pressure difference between hadronic and quark
phases due to the surface tension is neglected in the CP method.

\subsection{Gibbs construction}
\label{sec:4-3}

In the Gibbs construction, the finite-size effects due to surface and Coulomb
contributions are neglected completely, so the mixed phase does not include any pasta structures.
In this case, the surface tension at the hadron-quark interface is assumed to be negligible,
and global charge neutrality is required. Both hadronic matter and quark matter
are allowed to be charged separately.
Since only bulk contributions are considered, the energy density of the
mixed phase is reduced to
\begin{equation}
\varepsilon_{\rm{MP}} = u \varepsilon_{\rm{QP}}
  + (1 - u)\varepsilon_{\rm{HP}},
\label{eq:e3}
\end{equation}
where the surface and Coulomb terms vanish compared with Eq.~(\ref{eq:fws}).
The equilibrium conditions can be derived from the minimization of Eq.~(\ref{eq:e3}),
which have been given by Eqs.~(\ref{eq:CP2})--(\ref{eq:CE2}).
The pressure equilibrium between hadronic and quark phases is shown in Eq.~(\ref{eq:CP2}),
while Eq.~(\ref{eq:CU2}) represents the chemical potential equilibrium between
two phases. At a given baryon density $n_b$, there are two independent chemical potentials,
$\mu_{n}$ and $\mu_{e}$, which can be determined by the constraints of global charge
neutrality and baryon number conservation given in Eqs.~(\ref{eq:nc}) and~(\ref{eq:nb}).
The Gibbs equilibrium conditions of Eqs.~(\ref{eq:CP2}) and (\ref{eq:CU2}) imply that
coexisting hadronic and quark phases have equal pressures and chemical potentials.
After the equilibrium state is determined by Gibbs conditions,
all properties of the mixed phase can be calculated.

\subsection{Maxwell construction}
\label{sec:4-4}

In the Maxwell construction, the system satisfies the local charge neutrality condition,
namely, both hadronic and quark phases are charge neutral.
This is related to an extremely large surface tension at the hadron-quark interface,
which disfavors the formation of charged cluster of quark matter immersed in hadronic matter.
The energy density of the mixed phase includes only bulk contributions as described
in Eq.~(\ref{eq:e3}). Due to the local charge neutrality condition, there are three
constraints instead of Eqs.~(\ref{eq:nc}) and~(\ref{eq:nb}), which are expressed as
\begin{eqnarray}
0 &=& n_c^{\rm{HP}}=n_e^{\rm{HP}}+n_\mu^{\rm{HP}}-n_p  , \label{eq:nchp}\\
0 &=& n_c^{\rm{QP}}=n_e^{\rm{QP}}+n_\mu^{\rm{QP}}-\frac{1}{3}\left( 2n_u - n_d - n_s \right),\label{eq:ncqp}\\
n_b &=& \frac{u}{3}\left( n_u + n_d + n_s \right)
    + (1 - u) \left( n_p + n_n \right). \label{eq:nbm}
\end{eqnarray}
In the Maxwell construction, the electron density is usually discontinuous across the interface
due to local charge neutrality. We introduce the Lagrange multipliers, $\mu_e^{\rm{HP}}$,
$\mu_e^{\rm{QP}}$, and $\mu_n$, for these three constraints and construct a function as
\begin{eqnarray}
w &=&\varepsilon_{\rm{MP}}
  -\mu_n \left[ \frac{u}{3}\left( n_u + n_d + n_s \right)
    + (1 - u) \left( n_p + n_n \right)\right] \notag \\
& & -\mu_e^{\rm{QP}} u \left[n_e^{\rm{QP}}+n_\mu^{\rm{QP}}-\frac{1}{3}\left(2n_u-n_d-n_s\right)\right]\notag \\
& & -\mu_e^{\rm{HP}} (1 - u) \left[ n_e^{\rm{HP}}+n_\mu^{\rm{HP}}-n_p  \right].
\end{eqnarray}
By minimizing $w$ with respect to the volume fraction $u$ and particle densities,
we obtain the Maxwell conditions for phase equilibrium:
\begin{eqnarray}
P_{\rm{HP}} &=& P_{\rm{QP}}, \label{eq:CP4} \\
\mu_n &=& \mu_u+2\mu_d. \label{eq:CU4}
\end{eqnarray}
Meanwhile, the $\beta$ equilibrium conditions in hadronic and quark matter are respectively expressed as
\begin{eqnarray}
\mu_p+\mu_e^{\rm{HP}} &=& \mu_n, \\
\mu_u+\mu_e^{\rm{QP}} &=& \mu_d = \mu_s.
\end{eqnarray}
The Maxwell equilibrium conditions mean that coexisting hadronic and quark phases
have the same pressure and baryon chemical potential but different electron chemical potential.
During the phase transition, the pressure of the mixed phase in the Maxwell construction
remains constant. This behavior is different from that in the Gibbs construction,
where the pressure of the mixed phase increases with increasing density.

\section{Results and discussion}
\label{sec:5}

In this section, we investigate and compare the pasta structures of hadron-quark
mixed phase using the methods described in the previous section.
In order to check the model dependence of the results, we use two different
RMF models (TM1 and IUFSU) for the description of hadronic matter.
Meanwhile, the effects of repulsive vector interactions in the NJL model
are also examined. By employing a set of models with different symmetry energy
slope $L$, we discuss the effects of nuclear symmetry energy on the hadron-quark
phase transition. The properties of neutron stars are calculated by using the EOS
including quark degrees of freedom.

\subsection{Pasta structures in hadron-quark mixed phase}
\label{sec:5-1}

During the hadron-quark phase transition, several pasta configurations may
appear in the order of droplet, rod, slab, tube, and bubble with increasing density.
It is interesting to check whether all these geometric shapes would occur in the
mixed phase and how the pasta phases are affected by the model parameters.
To study the pasta structures in hadron-quark mixed phase, we employ the EM and CP
methods described in Sec.~\ref{sec:4}. It is known that the geometric
shape and size of the mixed phase are mainly determined by competition between
surface and Coulomb energies. Therefore, only the EM and CP methods can be used,
whereas the Gibbs and Maxwell constructions cannot describe pasta structures
due to the absence of surface and Coulomb contributions.
In the EM method, the equilibrium conditions between two coexisting phases
are determined by minimizing the total energy including surface and Coulomb
contributions. However, in the CP method, the Gibbs equilibrium conditions are
adopted that correspond to the balance without finite-size effects, while the
surface and Coulomb energies are perturbatively incorporated after the equilibrium
state is achieved.

\begin{figure}[htbp]
\includegraphics[bb=55 45 580 580, width=7 cm,clip]{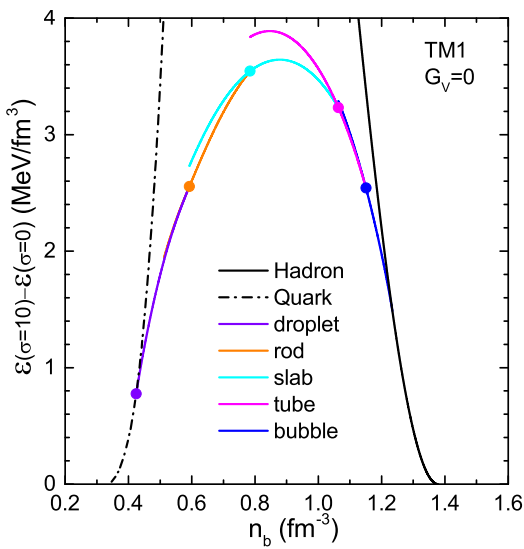}
\caption{(Color online) Comparison of energy densities for various pasta phases
obtained using the EM method with $\sigma=10$ MeV/fm$^2$ and $G_V=0$
relative to those of the Gibbs construction ($\sigma=0$).
The filled circles indicate the transition points between different configurations.}
\label{fig:1enb}
\end{figure}

Generally, the energy density difference between two successive configurations
is very small compared with the total energy density. In Fig.~\ref{fig:1enb},
we compare the energy densities of various pasta phases for
$\sigma = 10$ MeV/fm$^2$ obtained using the EM method relative to those of
the Gibbs construction ($\sigma = 0$).
The results are calculated with the TM1 parametrization given
in Table~\ref{tab:para}, while the vector coupling ${G_V}=0$ is adopted in the
NJL model. For comparison, the energy densities of pure hadronic and pure
quark phases are plotted by black dot-dashed and solid lines, respectively.
At a given baryon density $n_b$, the equilibrium state is the one with
the lowest energy density. When the energy density of a droplet becomes lower than
that of pure hadronic matter, quark droplets are formed in hadronic matter and
the hadron-quark phase transition starts. As the density increases, other
pasta configurations, such as rod, slab, etc., may appear when each has the
lowest energy density among all configurations. The phase transition
ends at the point where pure quark matter has lower energy density than
pasta phases. It is seen in Fig.~\ref{fig:1enb} that the energy density
difference between two successive shapes is rather small, and it is almost
invisible between the droplet (bubble) and rod (tube) phases.

\begin{figure*}[htbp]
\includegraphics[bb=55 45 580 590, width=7 cm,clip]{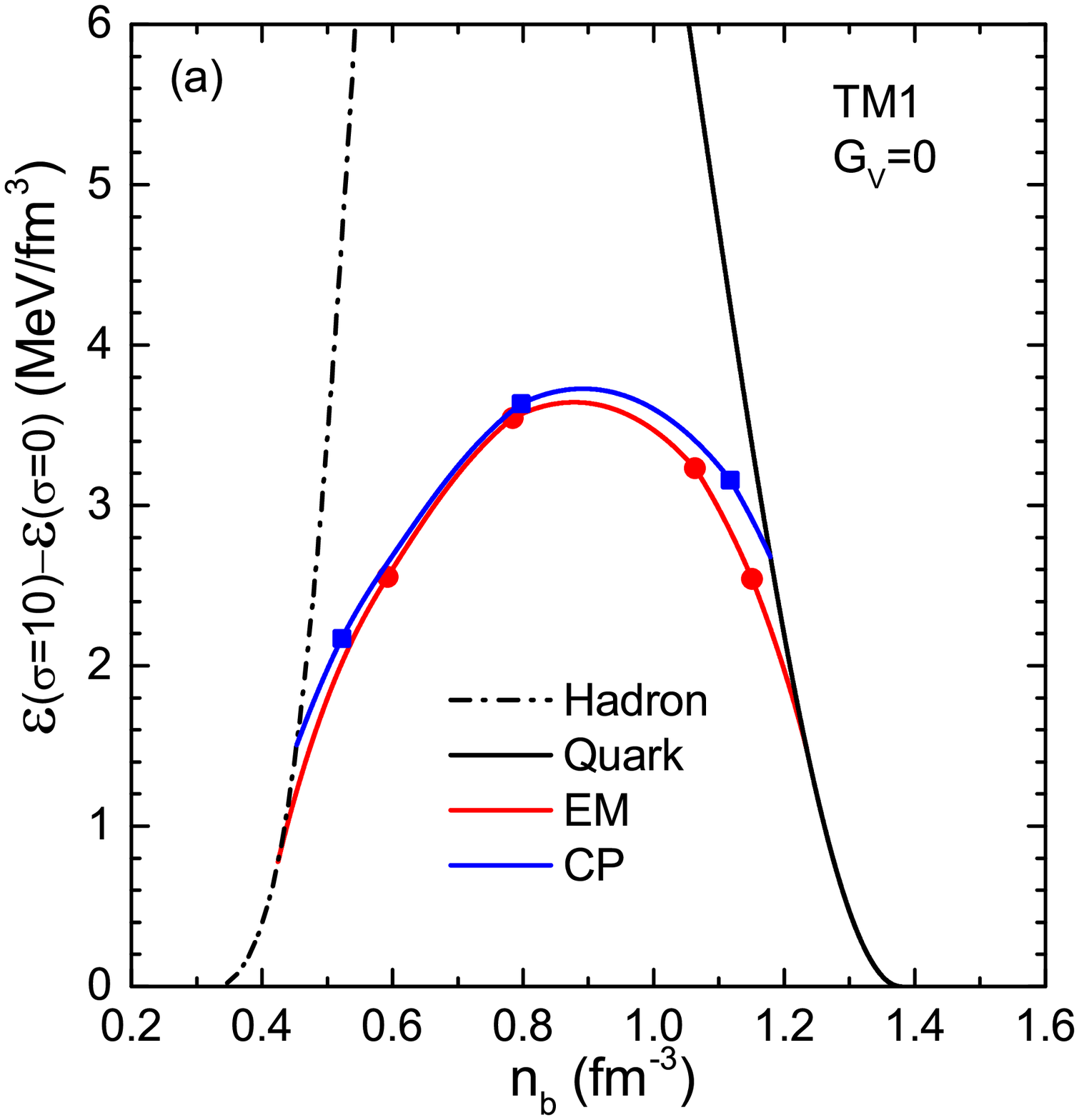}
\includegraphics[bb=55 45 580 590, width=7 cm,clip]{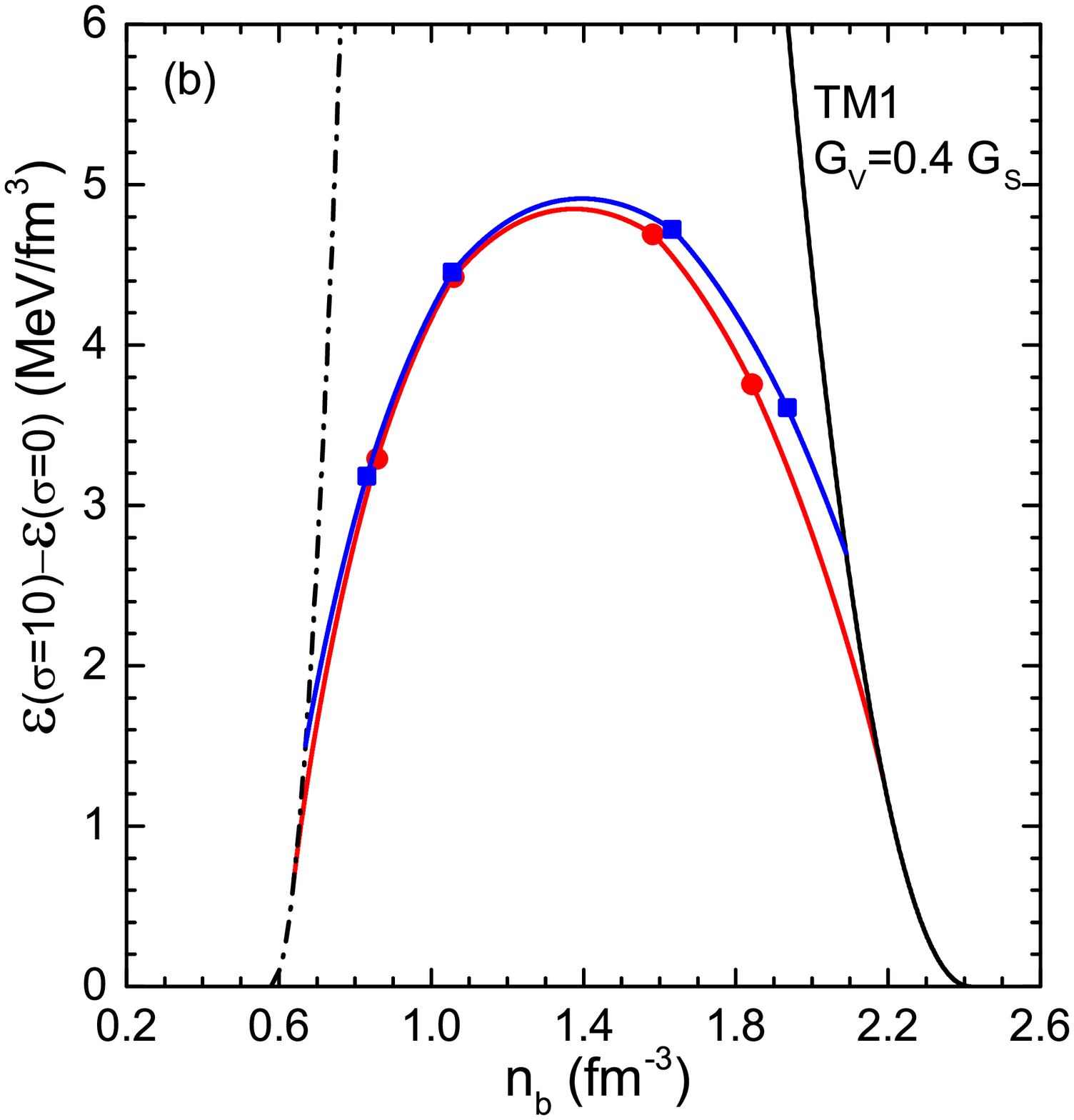}
\caption{(Color online) Comparison of energy densities obtained using the EM and CP methods
with $\sigma=10$ MeV/fm$^2$, relative to those of the Gibbs construction ($\sigma=0$).
The filled circles (squares) indicate the transition points between different configurations for EM (CP) method.}
\label{fig:2enb}
\end{figure*}

To evaluate the difference between the EM and CP methods, we compare in
Fig.~\ref{fig:2enb} the energy densities of pasta phases obtained from the
two methods with the surface tension $\sigma = 10$ MeV/fm$^2$, where the
transition points between different configurations are indicated by filled circles.
In the calculations, the TM1 model is used for hadronic phase,
while the NJL model with ${G_V}=0$ and ${G_V}=0.4\,G_S$ are
adopted for quark phase in the left and right panels, respectively.
It is shown that the energy densities of the EM method are slightly lower than
those of the CP method. This is because there are relatively large
configuration space in the EM method, and therefore lower energies could
be achieved in the minimization procedure.
By comparing the two panels of Fig.~\ref{fig:2enb}, one can see that
the energy densities for ${G_V}=0.4\,G_S$ are significantly larger than
those for ${G_V}=0$, and the density range of the mixed phase for ${G_V}=0.4\,G_S$
is shifted to larger values. This is because repulsive vector interactions
in the NJL model can effectively stiffen the EOS of quark matter, which
results in a delay of the phase transition.

\begin{figure*}[htbp]
\includegraphics[bb=50 5 580 400, width=0.31 \linewidth,clip]{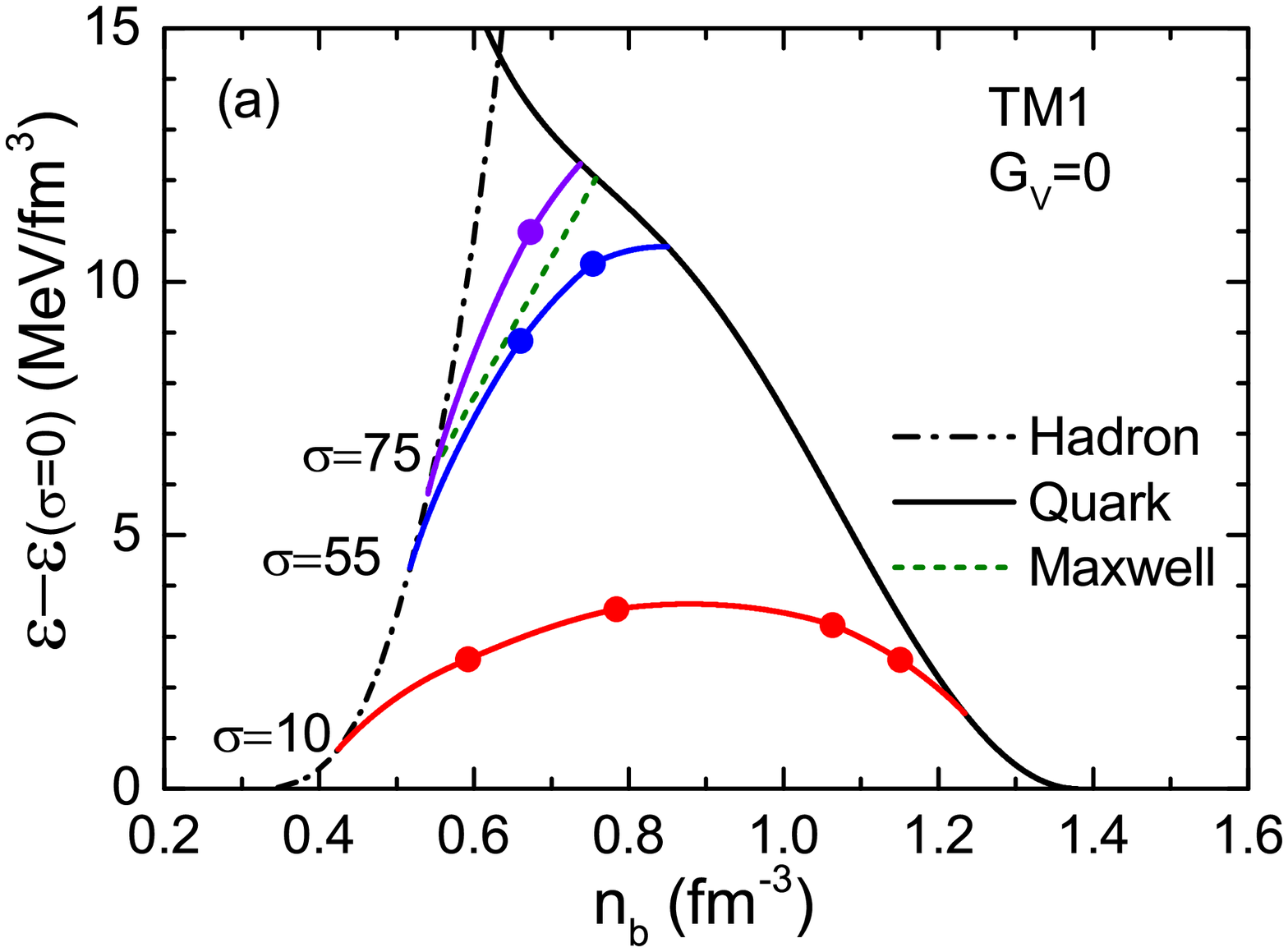}
\includegraphics[bb=50 5 580 400, width=0.31 \linewidth,clip]{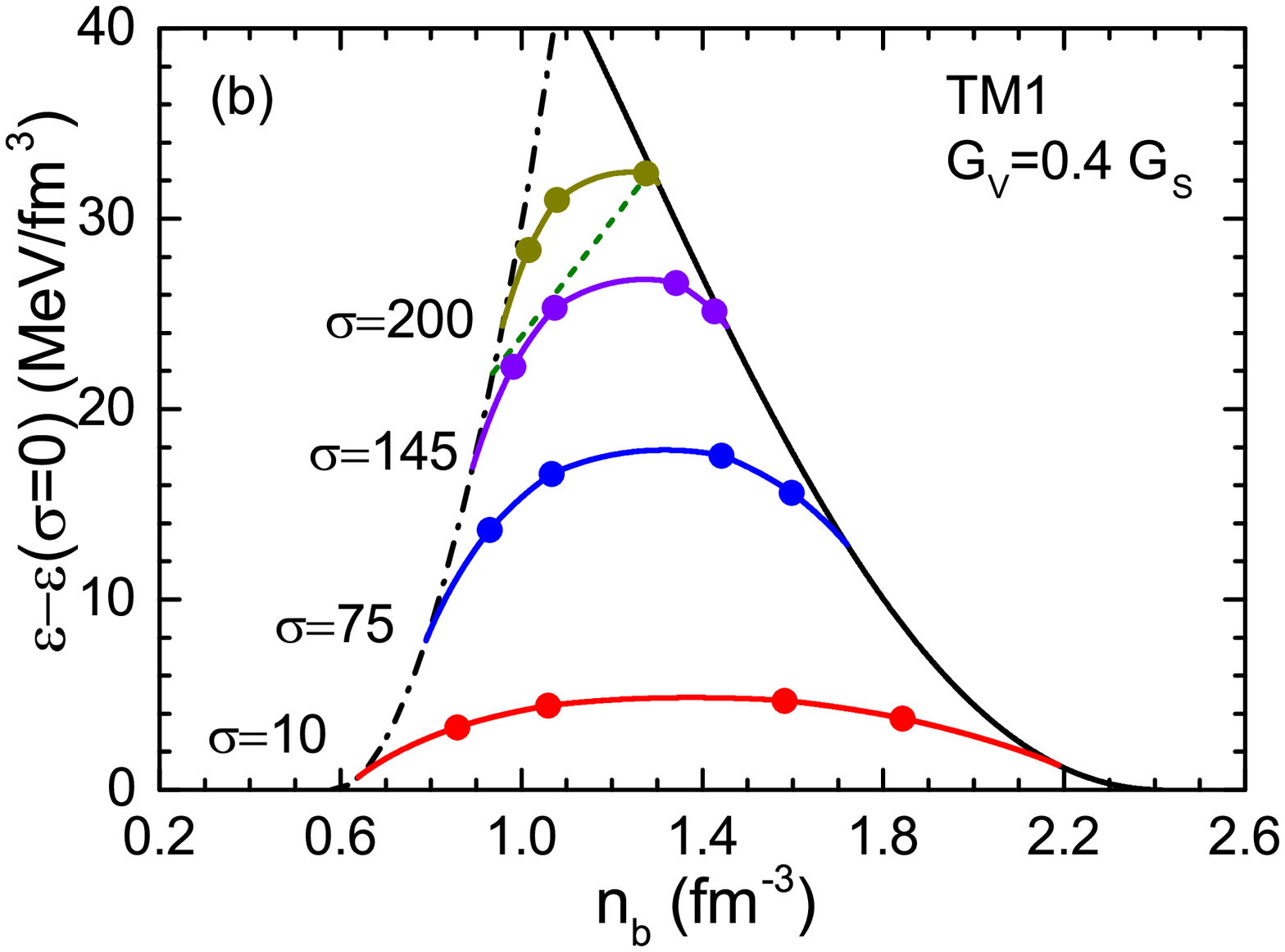}
\includegraphics[bb=50 5 580 400, width=0.31 \linewidth,clip]{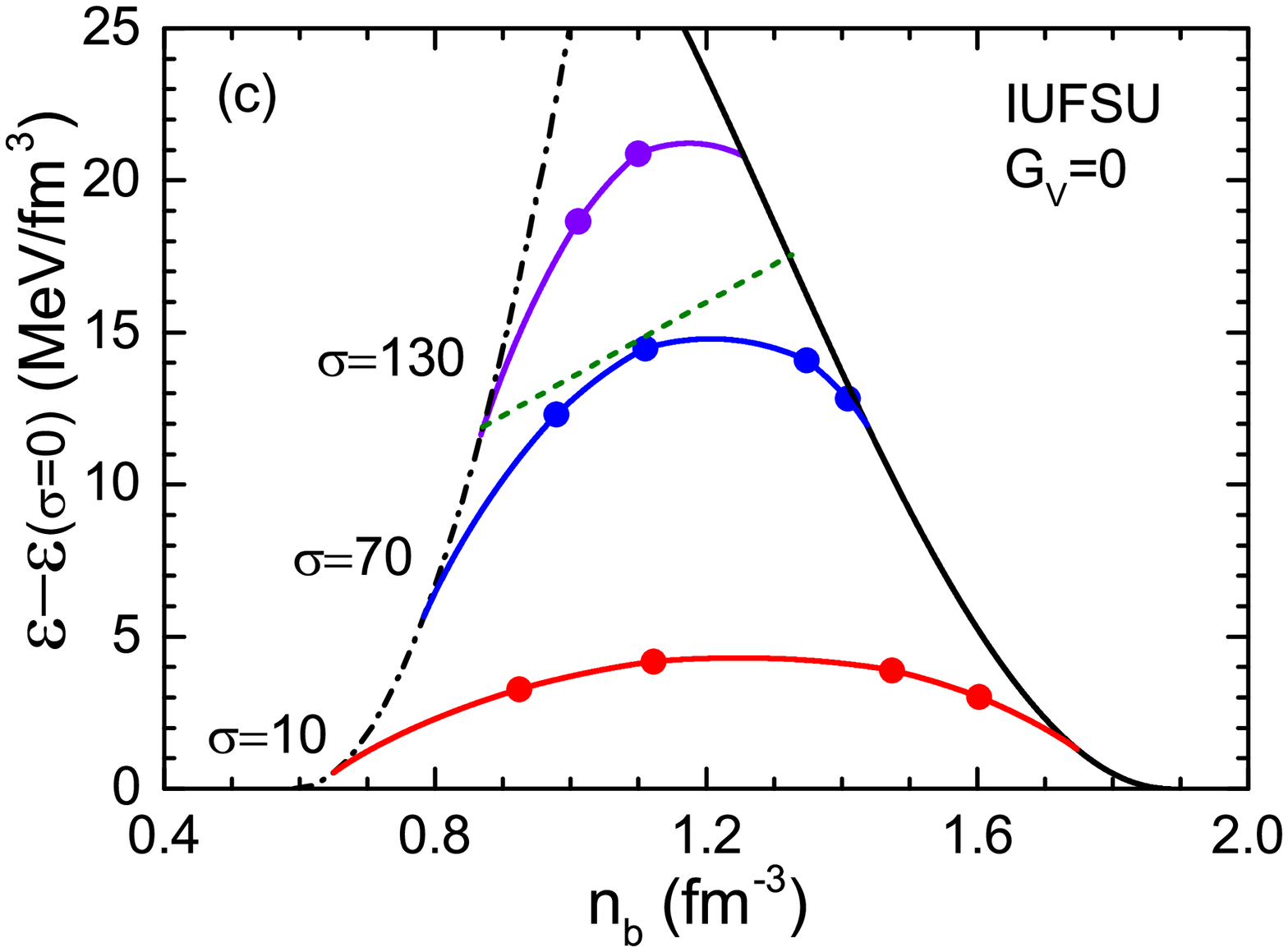}
\caption{(Color online) Energy densities of the mixed phase obtained using the
EM method for several values of $\sigma$ relative to those of the Gibbs construction ($\sigma=0$).
The filled circles indicate the transition points between different configurations.
The results of the Maxwell construction are shown by the green dotted lines.}
\label{fig:3enb}
\end{figure*}

In order to study how the surface tension $\sigma$ affects the properties of
pasta phases, we present in Fig.~\ref{fig:3enb} the energy densities of the
mixed phase obtained using the EM method for several values of $\sigma$
relative to those of the Gibbs construction ($\sigma = 0$).
The filled circles indicate the transition points between different configurations.
For comparison, the results obtained in the Maxwell construction are shown
by green dotted lines.
In the left panel of Fig.~\ref{fig:3enb}, the results are obtained by using the TM1 model
for hadronic phase and the NJL model with ${G_V}=0$ for quark phase.
One can see that a larger value of $\sigma$ leads to a smaller density range
and less pasta structures in the mixed phase. With $\sigma=55$ MeV/fm$^2$,
only droplet, rod, and slab configurations can occur before the system turns
to pure quark matter.
When the surface tension is larger than the critical value of $\approx 75$ MeV/fm$^2$,
the pasta phase is energetically unfavorable because its energy density is higher than
that of the Maxwell construction. This means that the Maxwell construction
is preferred for such high surface tension. It is noticeable that no mixed
phase would occur inside neutron stars for $\sigma > 75$ MeV/fm$^2$ because
the energetically favored Maxwell construction corresponds to constant pressure.
By comparing the left panel (${G_V}=0$) with the middle panel ($G_V=0.4\,G_S$)
in Fig.~\ref{fig:3enb}, we see that the repulsive vector interactions in the NJL
model can significantly shift the mixed phase toward higher densities with a wider
range. This is because the inclusion of repulsive vector interactions increases
the energy density of quark matter considerably. Meanwhile,
the critical surface tension increases from $\approx 75$ MeV/fm$^2$ for $G_V=0$
to $\approx 200$ MeV/fm$^2$ for $G_V=0.4\,G_S$ in the TM1 model.
The results of the right panel in Fig.~\ref{fig:3enb} correspond to the case
where the IUFSU model is used for hadronic phase and the NJL model with ${G_V}=0$
for quark phase. It is found that the mixed phase in the IUFSU model (right panel)
is shifted to higher densities with a wider range than that in the TM1
model (left panel), and meanwhile the critical surface tension required by
the Maxwell construction rises to $\approx 130$ MeV/fm$^2$ in the IUFSU model
from $\approx 75$ MeV/fm$^2$ in the TM1 model.
This is mainly because the symmetry energy slope $L$ in the IUFSU model
is much smaller than the one in the TM1 model.
The effects of the symmetry energy slope $L$ on the pasta phase properties
will be discussed in Sec.~\ref{sec:5-2}.

\begin{center}
\begin{figure*}[htb]
\centering
\includegraphics[bb=5 5 535 595, width=0.31 \linewidth,clip]{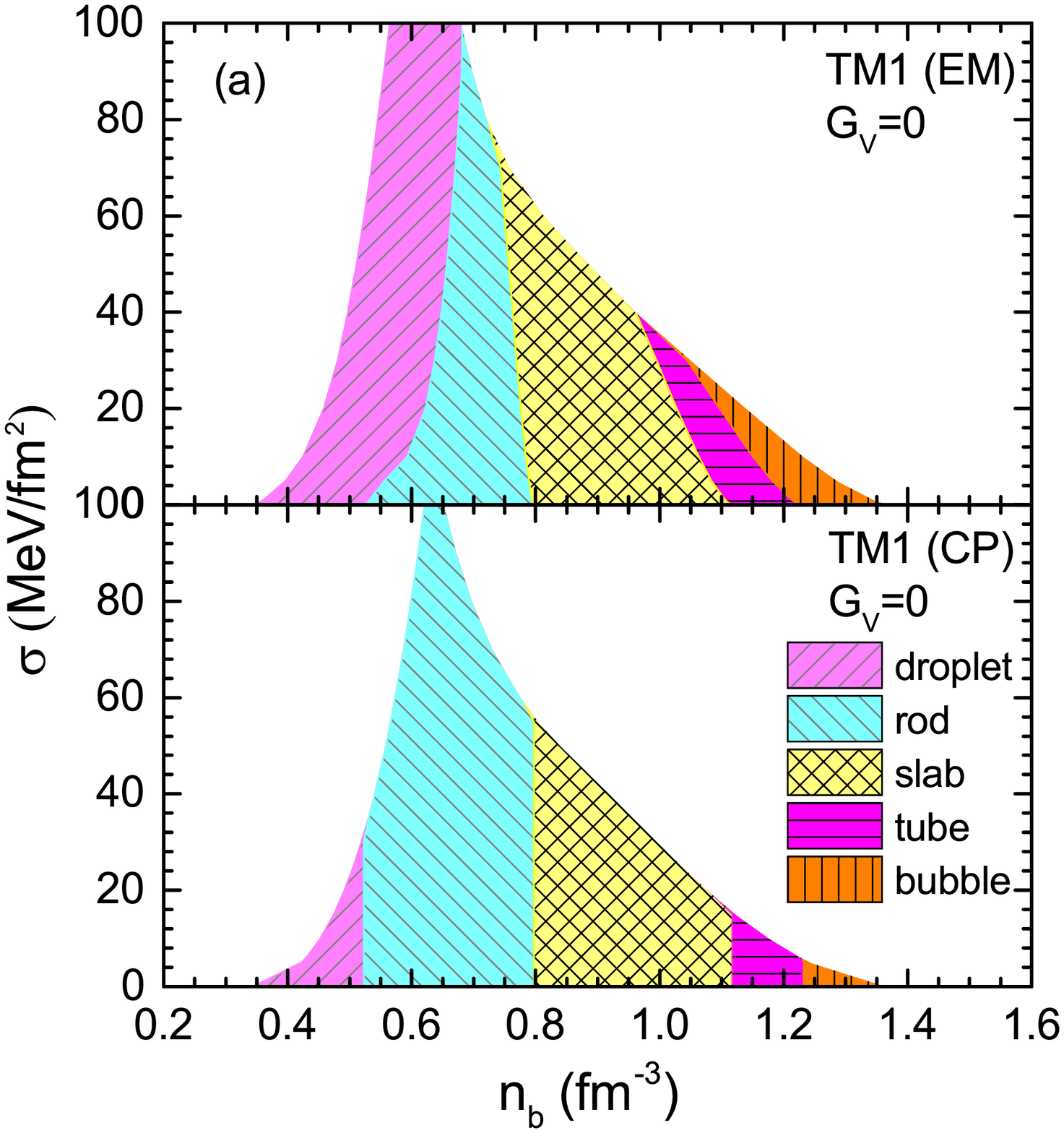}
\includegraphics[bb=5 5 535 595, width=0.31 \linewidth,clip]{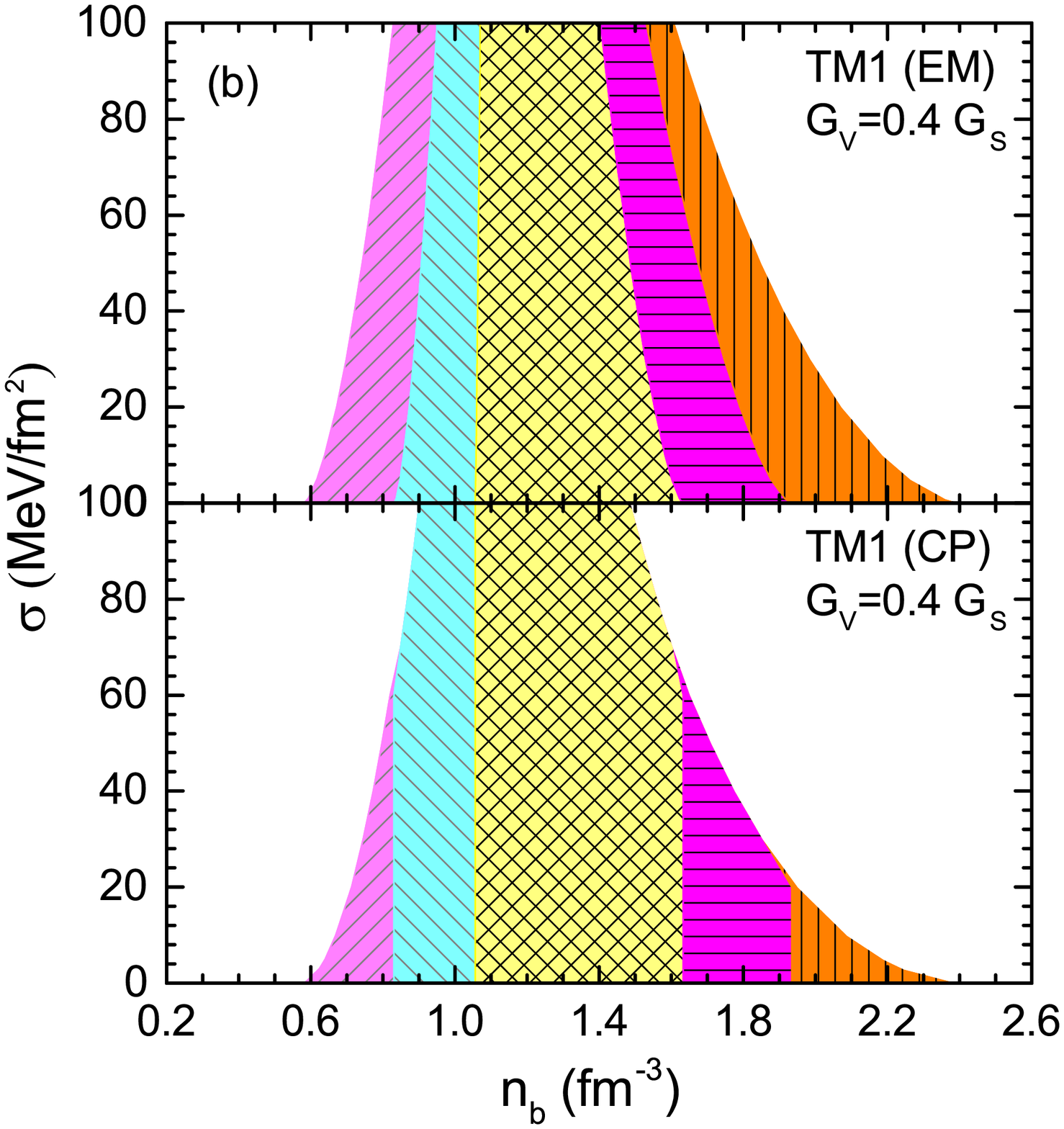}
\includegraphics[bb=5 5 535 595, width=0.31 \linewidth,clip]{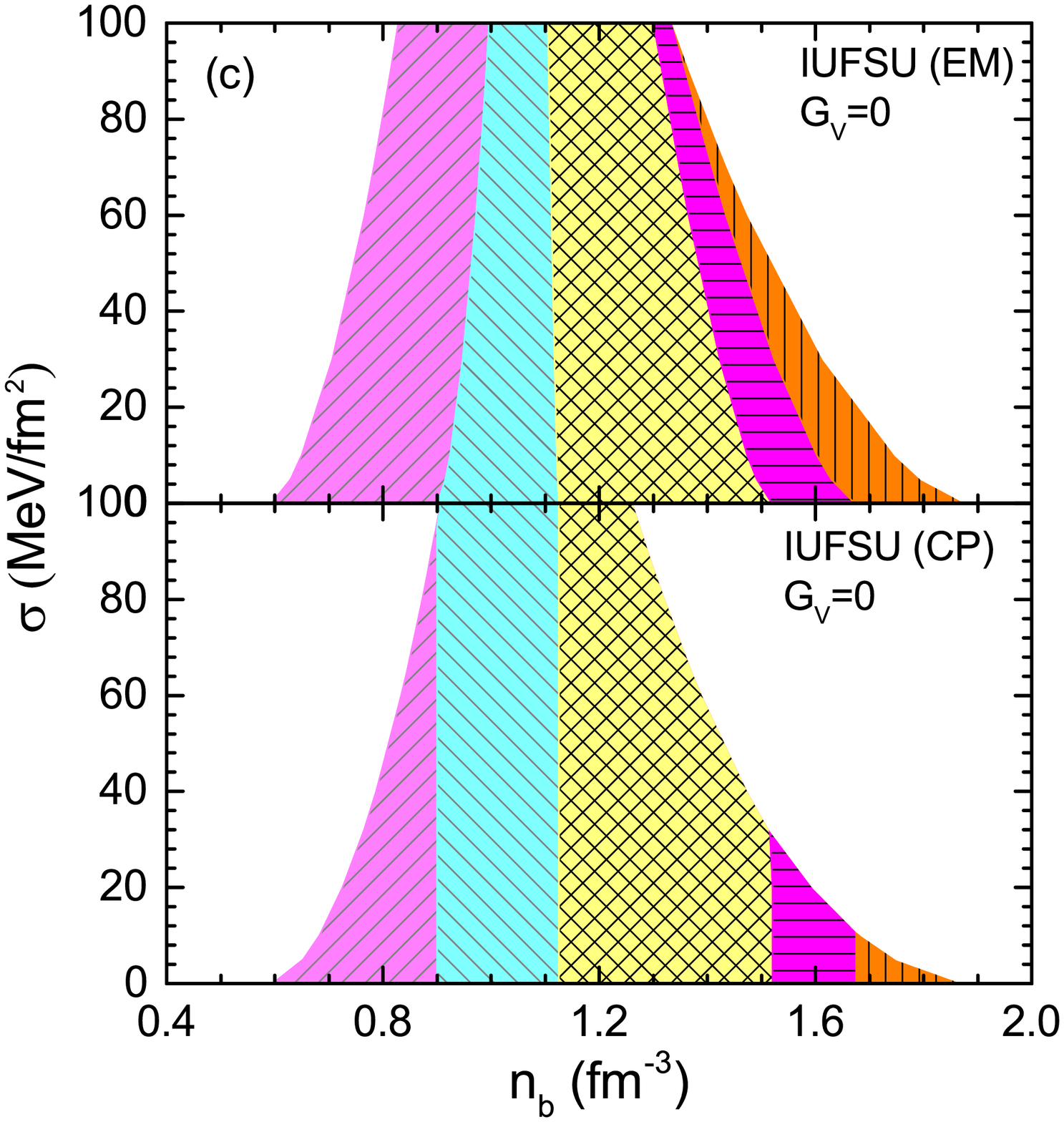}
\caption{(Color online) Density ranges of various pasta shapes as a function of the surface
tension $\sigma$. The results obtained using the EM and CP methods are displayed in
the upper and lower panels, respectively.}
\label{fig:4PT}
\end{figure*}
\end{center}

In Fig.~\ref{fig:4PT}, we show the density ranges of various pasta shapes
as a function of the surface tension $\sigma$. The results obtained from
the EM and CP methods are displayed in the upper and lower panels, respectively.
We can see that the onsets of all pasta shapes in the CP method are independent
of $\sigma$. This is because the equilibrium state in the CP method is determined
by the Gibbs conditions, which are unrelated to the surface tension $\sigma$.
At a given baryon density $n_b$, the favorable pasta shape is determined by
the sum of ${\varepsilon}_{\rm{surf}}+{\varepsilon}_{\rm{Coul}}$, which is
proportional to $\sigma^{2/3}$ derived from Eqs.~(\ref{eq:esurf})--(\ref{eq:rc}).
The transition between two pasta shapes occurs at the density where their
energy difference changes sign. Therefore, the transition density in the CP method
cannot be influenced by the surface tension $\sigma$ due to the simple
dependence ${\varepsilon}_{\rm{surf}}+{\varepsilon}_{\rm{Coul}}\propto \sigma^{2/3}$.
However, the dependence of ${\varepsilon}_{\rm{surf}}+{\varepsilon}_{\rm{Coul}}$
on the surface tension $\sigma$ is much more complicated in the EM method, since
the finite-size effects have been included in the equilibrium conditions.
Therefore, the transition density obtained in the EM method is clearly dependent
on $\sigma$ as shown in the upper panels of Fig.~\ref{fig:4PT}.
As $\sigma$ increases, the density range of hadron-quark mixed phase significantly
shrinks and the number of pasta configurations is reduced.
As shown in the upper-left panel of Fig.~\ref{fig:4PT}, the pasta phases can be
formed even if the surface tension is larger than the critical value
of $\approx 75$ MeV/fm$^2$ required by the Maxwell construction. However, the mixed
phase for $\sigma> 75$ MeV/fm$^2$ would not occur in neutron stars, because the
energetically favored Maxwell construction corresponds to constant pressure.
It is found that the qualitative behaviors of pasta structures are very similar
in all panels of Fig.~\ref{fig:4PT}, although there are quantitative differences.
In the present work, we focus on the study of pasta structures in hadron-quark
mixed phase, so a relatively small surface tension ($\sigma=10$ MeV/fm$^2$) will
be used in the following calculations.

\begin{figure*}[htbp]
\includegraphics[bb=30 30 580 590, width=7 cm,clip]{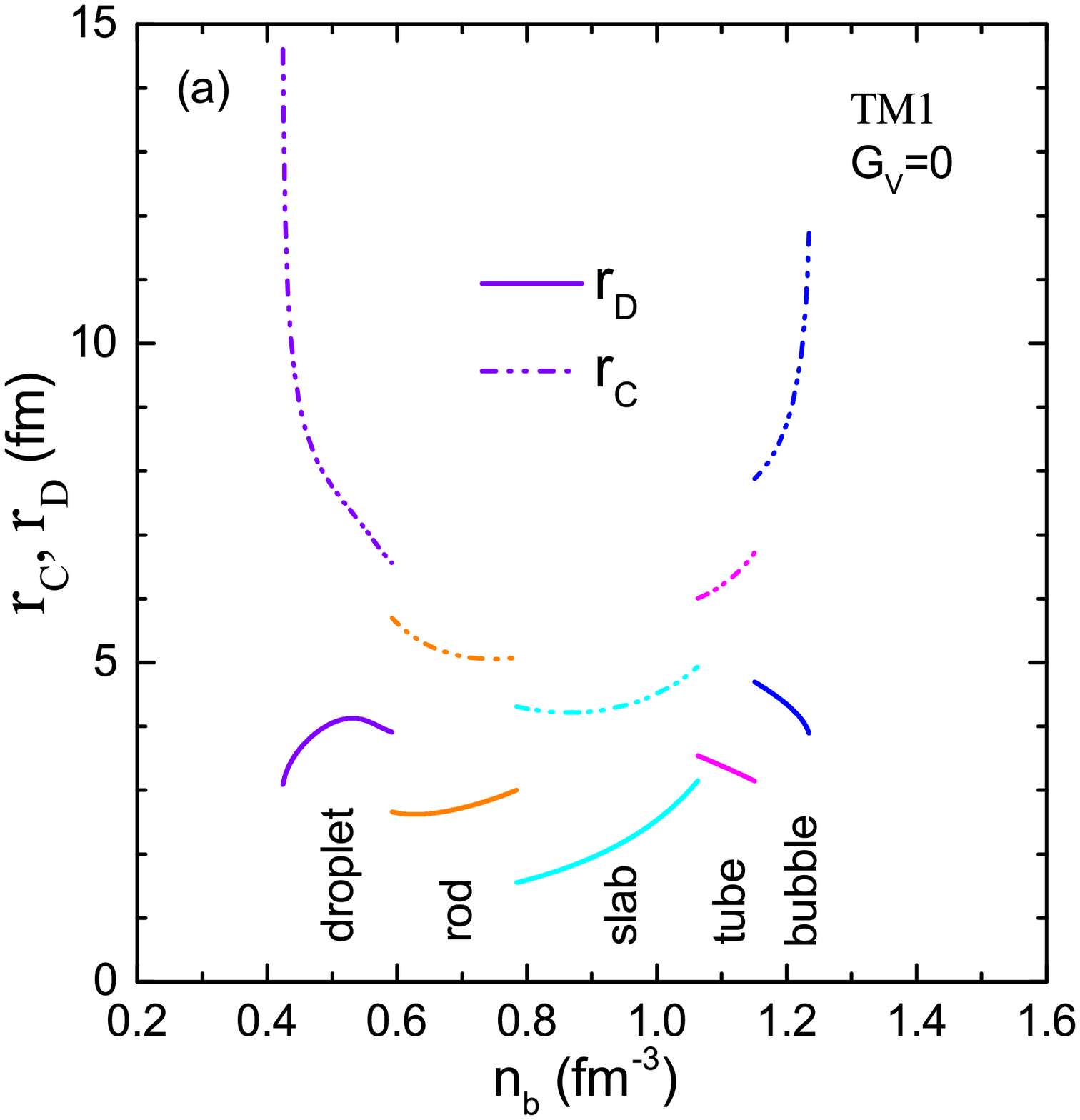}
\includegraphics[bb=30 30 580 590, width=7 cm,clip]{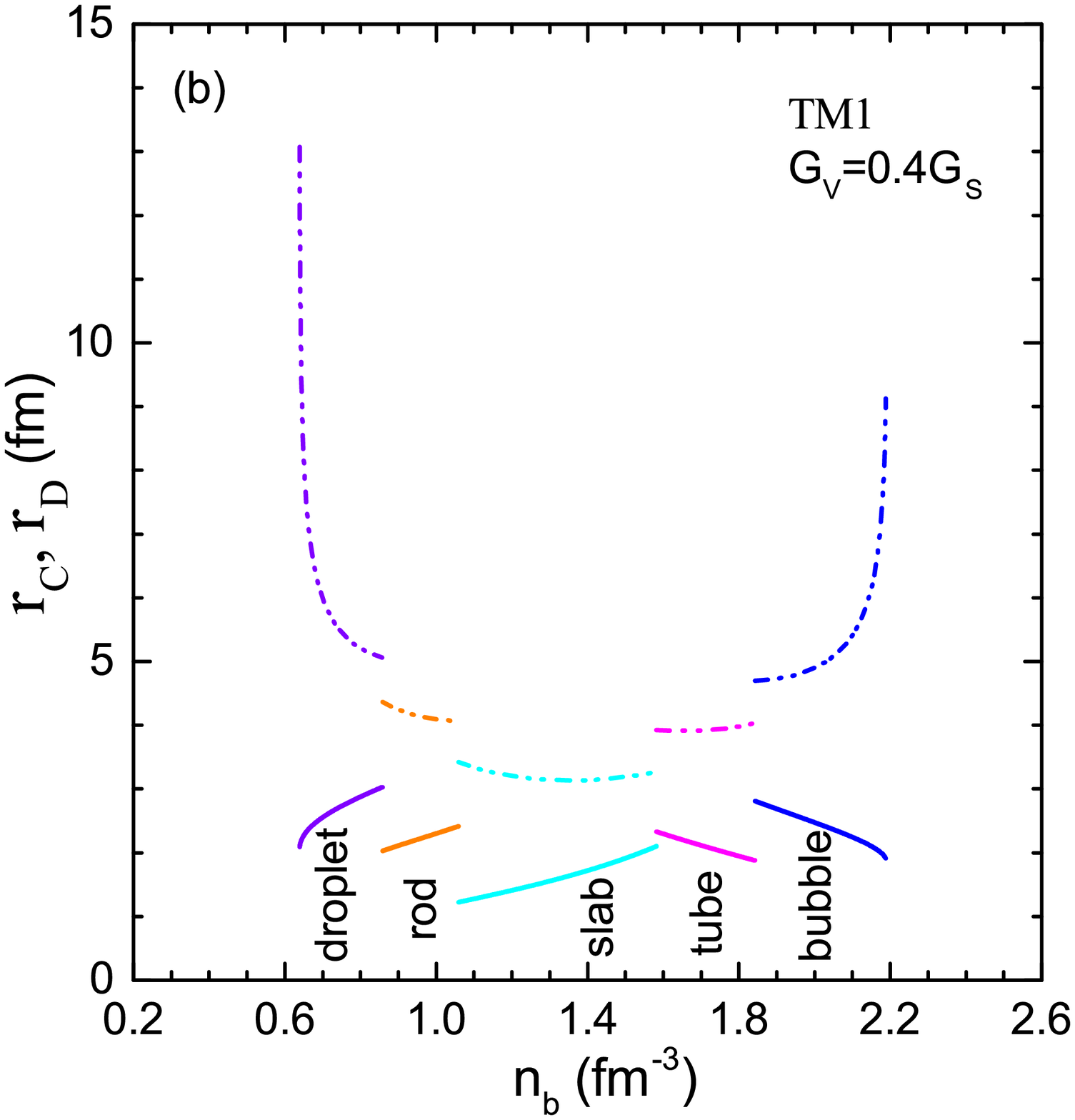}
\caption{(Color online) Size of the Wigner-Seitz cell ($r_C$) and that of the
inner part ($r_D$) as a function of $n_b$ obtained using the EM method
with $\sigma=10$ MeV/fm$^2$.}
\label{fig:5rnb}
\end{figure*}

In Fig.~\ref{fig:5rnb}, the size of the Wigner-Seitz cell ($r_C$) and that of
the inner part ($r_D$) obtained using the EM method in the TM1 model are displayed
as a function of the baryon density $n_b$. The results with ${G_V}=0$ and
${G_V}=0.4\,G_S$ are presented in the left and right panels, respectively.
It is found that there are obvious discontinuities in $r_D$ and $r_C$ at the
transition points between different shapes. One can see that $r_C$
decreases rapidly at lower densities, while it increases significantly in the bubble
phase before turning to pure quark matter. This behavior is related to
a monotonic increase of the volume fraction of quark phase, $u$, during the
phase transition. The tendency for ${G_V}=0.4\,G_S$ (right panel)
is similar to that for ${G_V}=0$ (left panel), but the density range
is shifted to larger values.

\subsection{Symmetry energy effects}
\label{sec:5-2}

\begin{figure*}[htbp]
\includegraphics[bb=30 70 590 590, width=12 cm,clip]{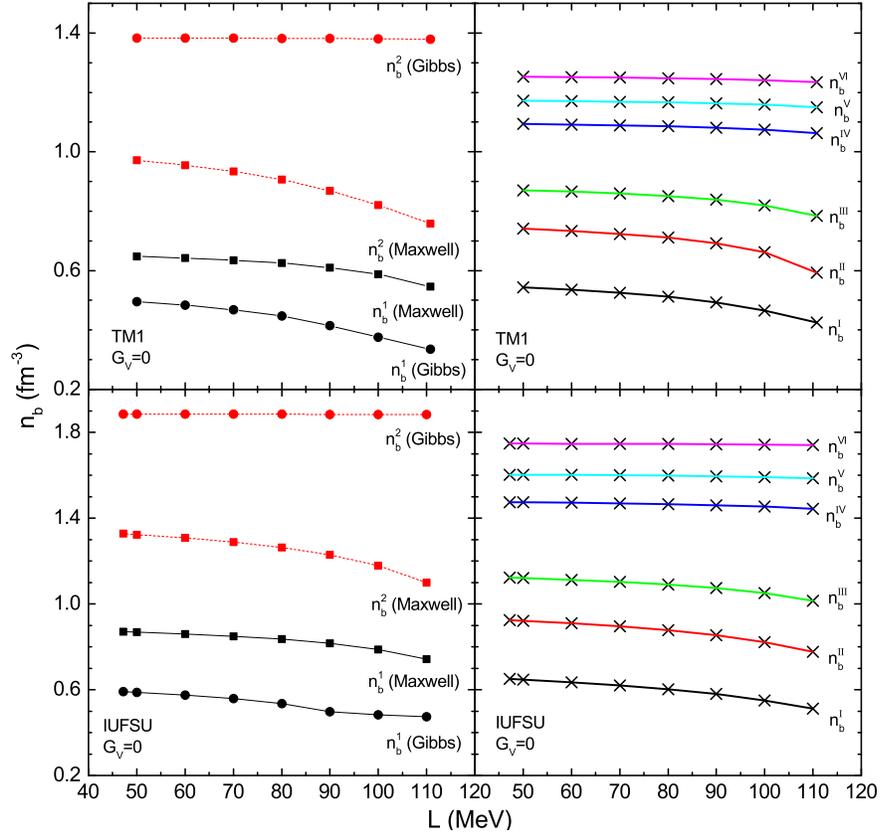}
\caption{(Color online) Transition densities as a function of the symmetry energy
slope $L$. The right panels show results obtained using the EM method
with $\sigma=10$ MeV/fm$^2$, where onset densities are in the order of
$n_b^{\rm{I}}$ (droplet), $n_b^{\rm{II}}$ (rod), $n_b^{\rm{III}}$ (slab),
$n_b^{\rm{IV}}$ (tube), $n_b^{\rm{V}}$ (bubble), and $n_b^{\rm{VI}}$ (pure quark matter).
The left panels present starting densities ($n_b^{1}$) and ending densities ($n_b^{2}$)
of the mixed phase obtained with the Gibbs and Maxwell constructions.}
\label{fig:6lnb}
\end{figure*}

To study the effects of nuclear symmetry energy on the hadron-quark phase
transition, we use two sets of generated RMF models based on the TM1 and IUFSU
parametrizations as described in Ref.~\cite{Bao14b}. We emphasize that all models
in each set have the same isoscalar saturation properties and fixed symmetry
energy $E_{\rm{sym}}$ at a density of $0.11\, \rm{fm}^{-3}$ but have different
symmetry energy slope $L$. Therefore, these models could predict very similar
properties of finite nuclei but different density dependence of nuclear symmetry
energy, which plays an important role in understanding the structure of neutron stars.
In Fig.~\ref{fig:6lnb}, we present the transition densities as a function of the
symmetry energy slope $L$ in the TM1 (upper panels) and IUFSU (lower panels) sets.
The results are obtained with ${G_V}=0$ and $\sigma = 10$ MeV/fm$^2$.
In the right panels, we display the onset densities of droplet ($n_b^{\rm{I}}$),
rod ($n_b^{\rm{II}}$), slab ($n_b^{\rm{III}}$), tube ($n_b^{\rm{IV}}$),
bubble ($n_b^{\rm{V}}$), and pure quark matter ($n_b^{\rm{VI}}$) obtained
using the EM method. In the left panels, we show starting densities ($n_b^{1}$)
and ending densities ($n_b^{2}$) of the mixed phase obtained with the Gibbs and
Maxwell constructions. Detailed results are also presented
in Table~\ref{tab:nt-L}.
One can see that as $L$ increases, all transition densities
decrease and the $L$ dependence becomes weaker at the end of the mixed phase.
This is because the fraction of hadronic matter monotonically decreases during
the hadron-quark phase transition, and therefore the influence of nuclear symmetry
energy gets weaker and weaker. It is shown that the onset densities of pure quark
matter, $n_b^{\rm{VI}}$ (right panels) and $n_b^{2}$ of Gibbs (left panels),
are almost independent of $L$.
\begin{table*}[htbp]
\caption{Onset densities of various phases obtained in the TM1 and IUFSU sets.
In the EM method, the surface tension $\sigma = 10$ MeV/fm$^2$ is used and onset densities
are in the order of $n_b^{\rm{I}}$ (droplet), $n_b^{\rm{II}}$ (rod), $n_b^{\rm{III}}$ (slab),
$n_b^{\rm{IV}}$ (tube), $n_b^{\rm{V}}$ (bubble), and $n_b^{\rm{VI}}$ (pure quark matter).
In the Gibbs and Maxwell constructions, $n_b^{1}$ and $n_b^{2}$
denote the starting and ending densities of the mixed phase, respectively.
All densities are in $\rm{fm}^{-3}$.}
\label{tab:nt-L}
\begin{center}
\setlength{\tabcolsep}{2mm}{
\begin{tabular}{ll|cc|cc|cccccccc}
\hline\hline
Model &  & \multicolumn{2}{c}{Gibbs} \vline&  \multicolumn{2}{c}{Maxwell} \vline& \multicolumn{6}{c}{EM}\\
\hline
 & $L$ (MeV) & $n_b^{1}$ & $n_b^{2}$ & $n_b^{1}$ & $n_b^{2}$ & $n_b^{\rm{I}}$ & $n_b^{\rm{II}}$ & $n_b^{\rm{III}}$  &$n_b^{\rm{IV}}$ &$n_b^{\rm{V}}$   & $n_b^{\rm{VI}}$   \\
\hline
TM1           &$ 50$  & 0.4957 & 1.3832   & 0.64774  & 0.97204 & 0.5438 & 0.7407      & 0.8708 & 1.0934 & 1.1720 & 1.2529\\
$G_{V}=0$     &$ 60$  & 0.4842 & 1.3828   & 0.64244  & 0.95529 & 0.5358 & 0.7333      & 0.8659 & 1.0915 & 1.1706 & 1.2516\\
              &$ 70$  & 0.4683 & 1.3825   & 0.63520  & 0.93349 & 0.5254 & 0.7236      & 0.8595 & 1.0891 & 1.1688 & 1.2500\\
              &$ 80$  & 0.4473 & 1.3820   & 0.62537  & 0.90600 & 0.5122 & 0.7109      & 0.8511 & 1.0861 & 1.1666 & 1.2479\\
              &$ 90$  & 0.4150 & 1.3813   & 0.61029  & 0.86855 & 0.4931 & 0.6918      & 0.8386 & 1.0816 & 1.1634 & 1.2451\\
              &$ 100$ & 0.3751 & 1.3803   & 0.58716  & 0.82095 & 0.4655 & 0.6626      & 0.8198 & 1.0751 & 1.1588 & 1.2410\\
            &$ 110.8$ & 0.3351 & 1.3786   & 0.54633  & 0.75782 & 0.4246 & 0.5924      & 0.7844 & 1.0632 & 1.1505 & 1.2340\\
\hline 
\tabincell{l}{TM1 \\ $G_{V}=0.4\,G_{S}$}          &$ 110.8$ & 0.5791 & 2.4156 & 0.93703 & 1.28827 & 0.6392 & 0.8577 & 1.0586 & 1.5825 & 1.8428 & 2.1879 \\
\hline
IUFSU         &$ 47.2$    &0.5914 & 1.8842      & 0.87046 & 1.32709  & 0.6499 & 0.9241      & 1.1218 & 1.4741 & 1.6025 & 1.7472\\
$G_{V}=0$     &$ 50$    &0.5880 & 1.8842      & 0.86836 & 1.32319  & 0.6468 & 0.9212      & 1.1199 & 1.4735 & 1.6022 & 1.7471\\
              &$ 60$    &0.5745 & 1.8841      & 0.85985 & 1.30751  & 0.6345 & 0.9097      & 1.1121 & 1.4712 & 1.6010 & 1.7466\\
              &$ 70$    &0.5579 & 1.8840      & 0.84920 & 1.28817  & 0.6199 & 0.8958      & 1.1025 & 1.4684 & 1.5995 & 1.7460\\
              &$ 80$    &0.5351 & 1.8838      & 0.83518 & 1.26304  & 0.6020 & 0.8781      & 1.0902 & 1.4649 & 1.5976 & 1.7452\\
              &$ 90$    &0.4972 & 1.8836      & 0.81574 & 1.22869  & 0.5795 & 0.8547      & 1.0738 & 1.4601 & 1.5952 & 1.7442\\
              &$ 100$   &0.4835 & 1.8833      & 0.78682 & 1.17803  & 0.5502 & 0.8221      & 1.0500 & 1.4532 & 1.5916 & 1.7428\\
              &$ 110$   &0.4733 & 1.8829      & 0.74264 & 1.09978  & 0.5111 & 0.7763      & 1.0144 & 1.4425 & 1.5916 & 1.7407\\
\hline\hline
\end{tabular}}
\end{center}
\end{table*}

\begin{figure*}[htb]
\includegraphics[bb=20 30 580 590, width=7 cm,clip]{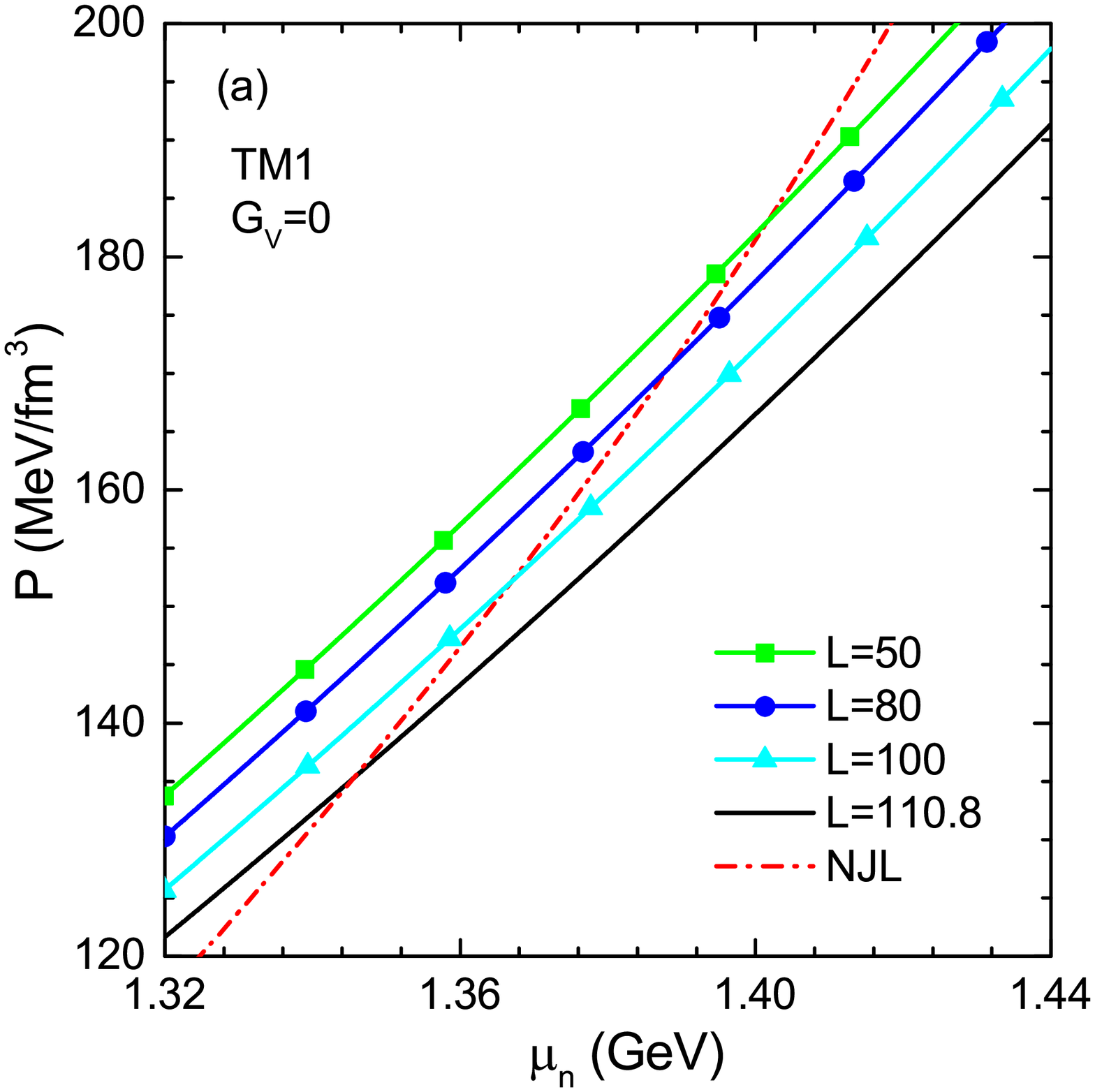}
\includegraphics[bb=20 30 580 590, width=7 cm,clip]{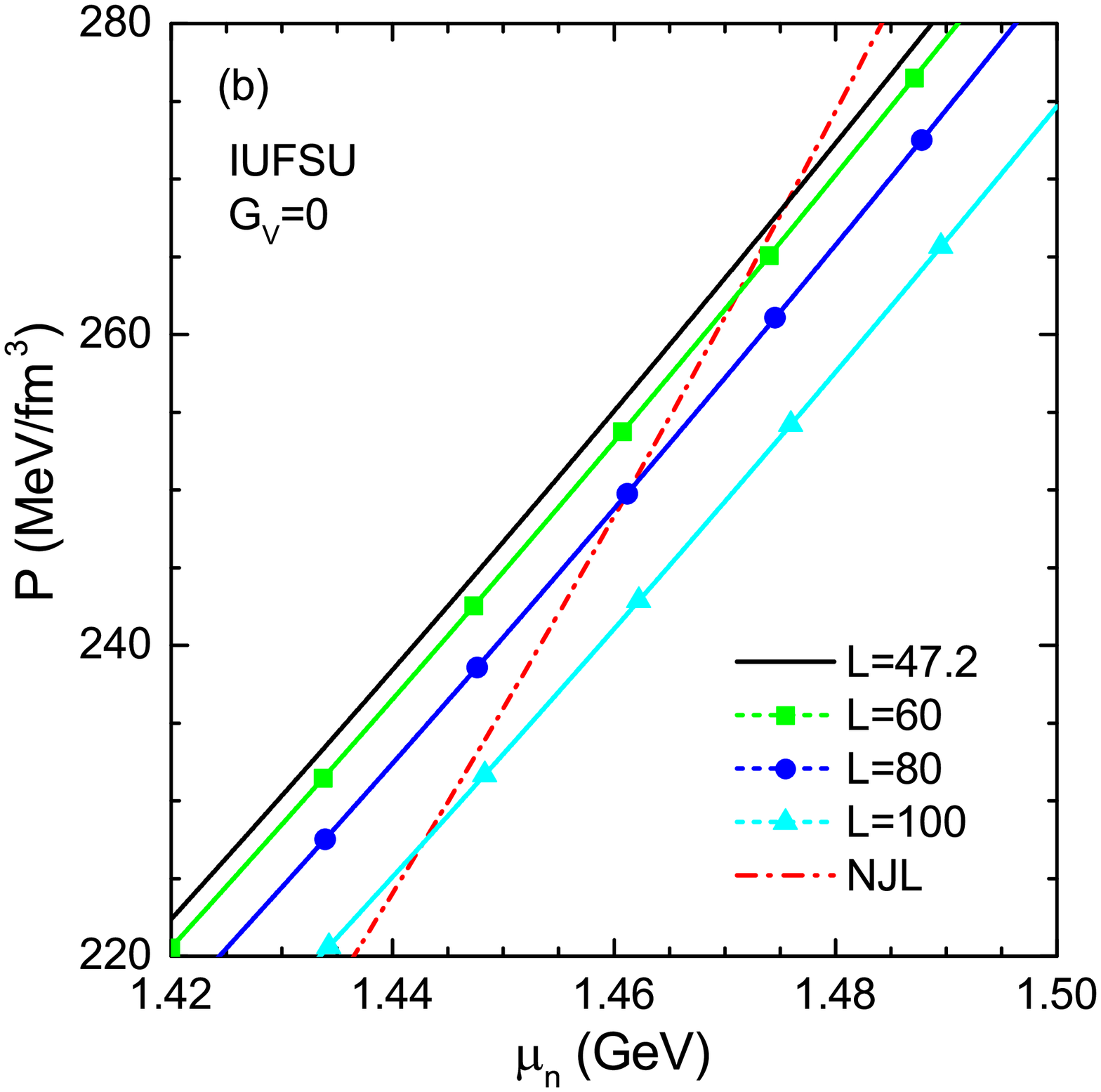}
\caption{(Color online) Pressure $P$ as a function of the neutron chemical
potential $\mu_n$ for different values of $L$.}
\label{fig:7xmunp}
\end{figure*}

In order to understand the $L$ dependence of the transition densities,
we show in Fig.~\ref{fig:7xmunp} the pressure $P$ as a function of the
neutron chemical potential $\mu_n$ for different values of $L$
in the TM1 (left panel) and IUFSU (right panel) sets.
According to the Maxwell equilibrium conditions given by Eqs.~(\ref{eq:CP4})
and (\ref{eq:CU4}), the phase transition occurs at the
crossing of the hadronic EOS with the quark EOS, where two phases have
the same pressure and neutron chemical potential.
In the hadronic phase, a smaller $L$ corresponds to a larger $P$,
which leads to a larger value of $\mu_n$ in the mixed phase with the Maxwell
constructions. Therefore, the transition densities for a small $L$
would be higher than those for a large $L$.

\begin{figure*}[htbp]
\includegraphics[bb=45 45 580 580, width=7 cm,clip]{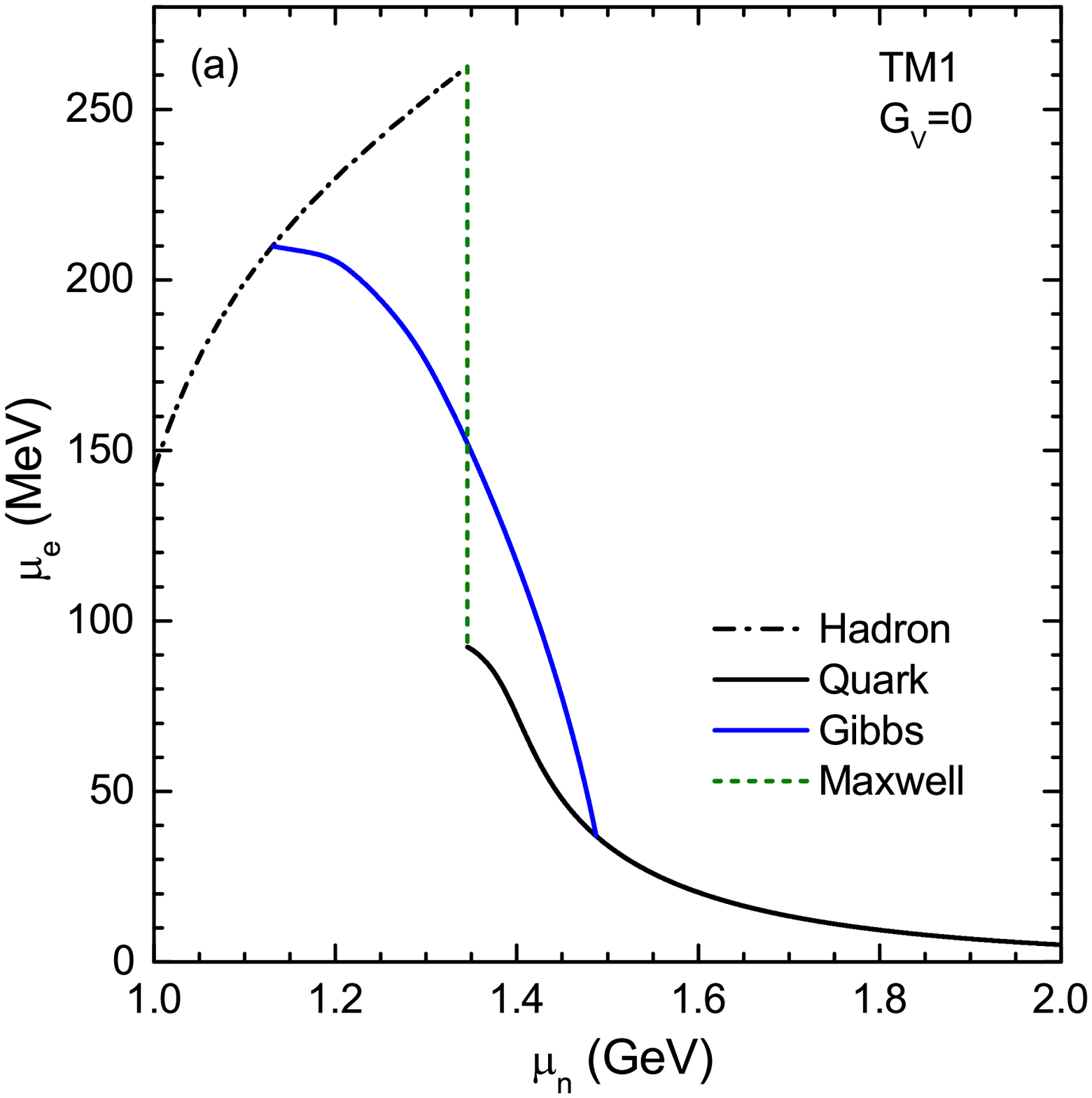}
\includegraphics[bb=45 45 580 580, width=7 cm,clip]{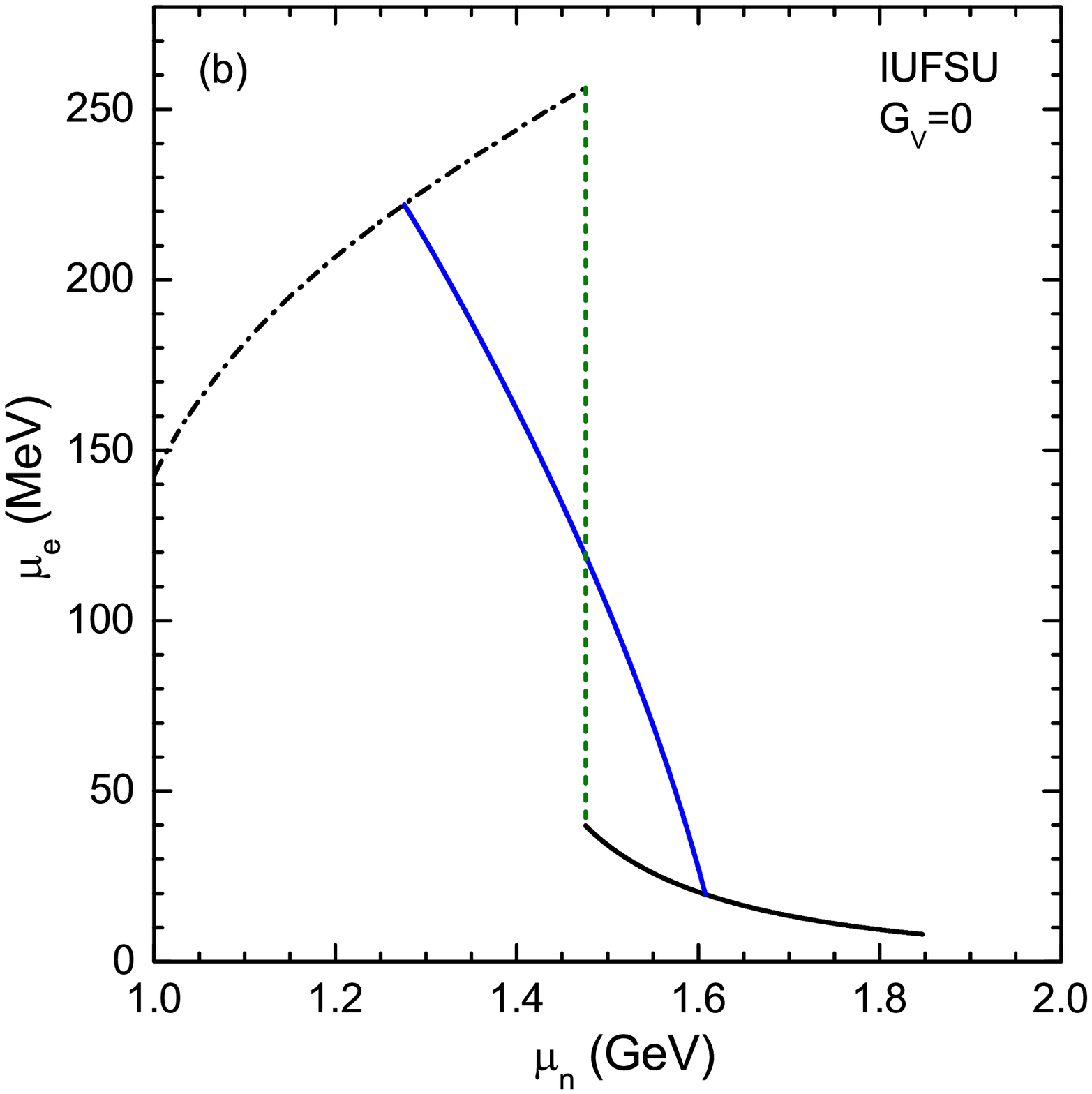}
\caption{(Color online) Electron chemical potential $\mu_e$ as a function of the neutron
chemical potential $\mu_n$ obtained with the Gibbs and Maxwell constructions.}
\label{fig:8xmunxmue}
\end{figure*}

It is interesting to look at the behavior of the electron chemical potential
$\mu_e$ and its $L$ dependence. Generally, $\mu_e$ is considered as a signal
of the imbalance between protons and neutrons in hadronic matter under $\beta$
equilibrium due to the relation $\mu_e={\mu_n}-{\mu_p}$, which is closely
related to nuclear symmetry energy $E_{\rm{sym}}$. Therefore, a larger $\mu_e$
implies that the system is more asymmetric.
In Fig.~\ref{fig:8xmunxmue}, we show the electron chemical potential $\mu_e$
as a function of the neutron chemical potential $\mu_n$ obtained with the Gibbs and
Maxwell constructions. The results of the original TM1 ($L=110.8$ MeV) and
IUFSU ($L=47.2$ MeV) models are presented in the left and right panels, respectively.
${G_V}=0$ is adopted in the NJL model.
It is seen that $\mu_e$ at the transition point with the Maxwell construction
is discontinuous, where $\mu_e$ of hadronic phase is much larger than that of
quark phase. By comparing the two panels of Fig.~\ref{fig:8xmunxmue},
we can see that $\mu_e$ of TM1 in pure hadronic matter is steeper than
that of IUFSU. This is because the symmetry energy slope $L$ of TM1 is
much larger than that of IUFSU. As a result, the TM1 model predicts larger
$\mu_e$ and smaller $\mu_n$ for the hadron-quark phase transition with
the Maxwell construction. We note that the pressure and chemical potentials
remain constant during the phase transition with the Maxwell construction.
However, for the Gibbs construction, $\mu_e$ and $\mu_n$ in the mixed phase
can extend over a finite range, and there is no abrupt jump in $\mu_e$ between
coexisting hadronic and quark phases. The behaviors of $\mu_e$ and $\mu_n$
obtained using the CP method should be the same as those of the Gibbs
construction, since the Gibbs conditions are used to determine the equilibrium
state in the CP method.

\subsection{Properties of neutron stars}
\label{sec:5-3}

\begin{figure*}[htbp]
\includegraphics[bb=30 5 580 420, width=7 cm,clip]{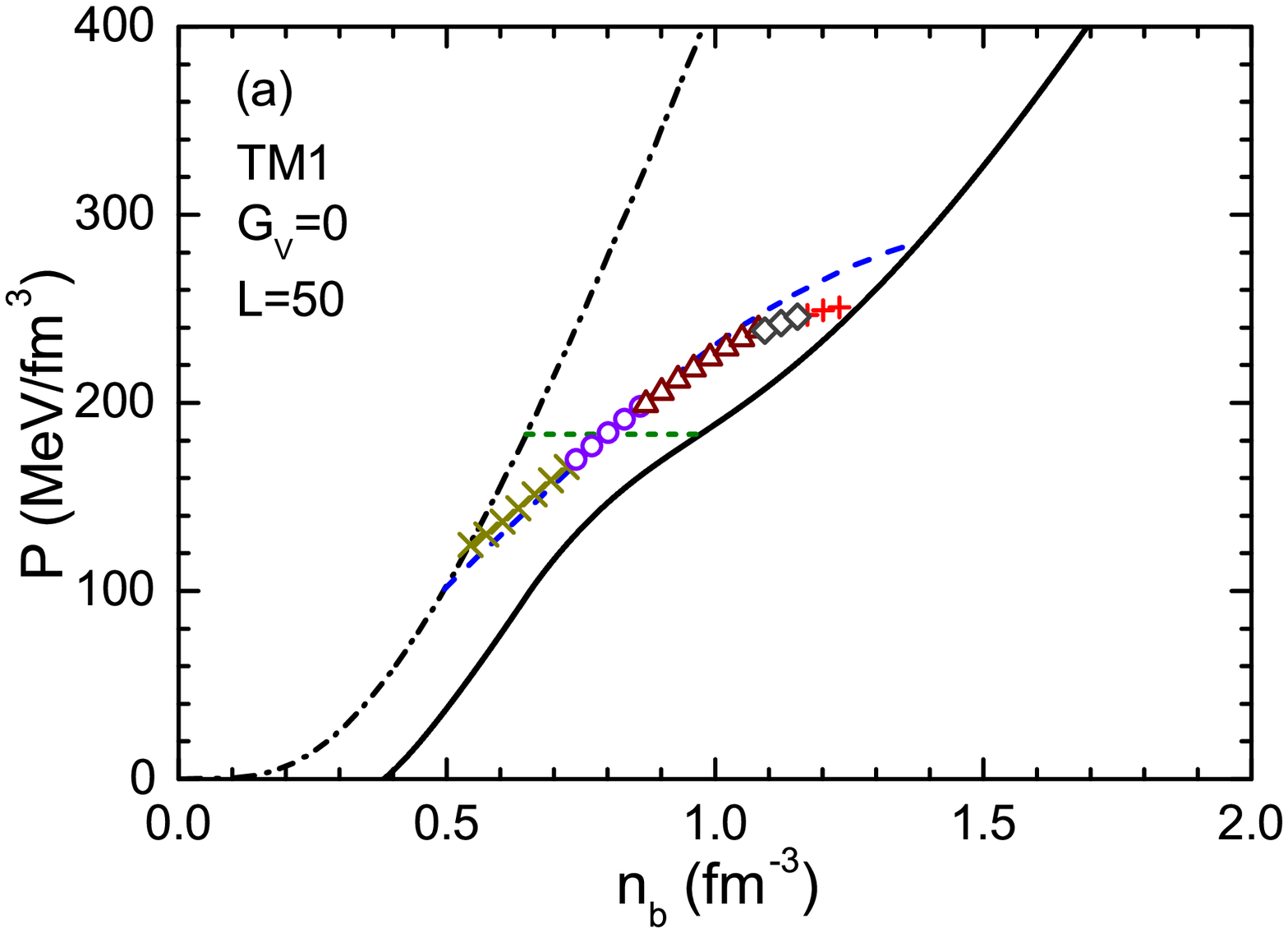}
\includegraphics[bb=30 5 580 420, width=7 cm,clip]{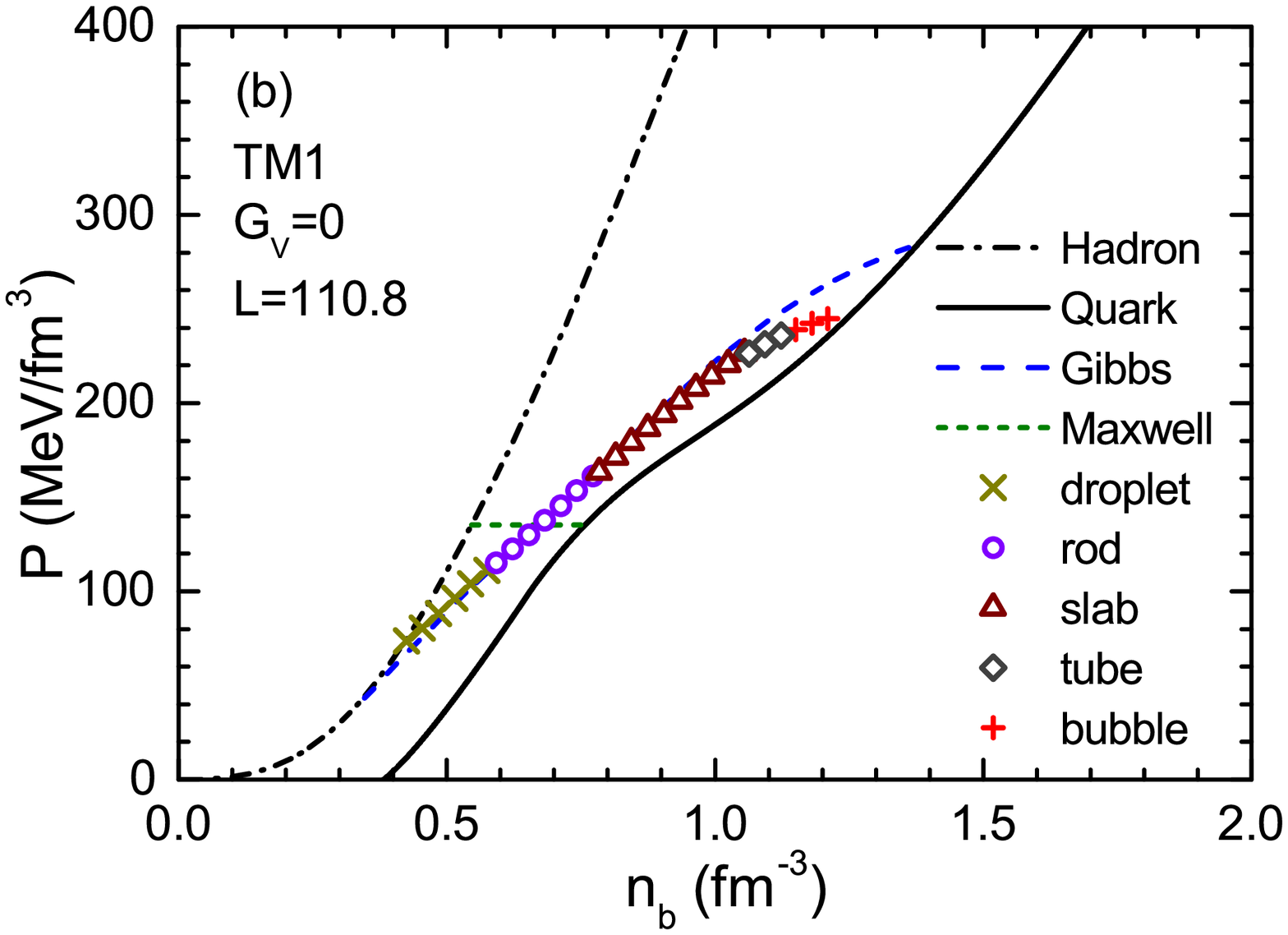}
\includegraphics[bb=30 5 580 420, width=7 cm,clip]{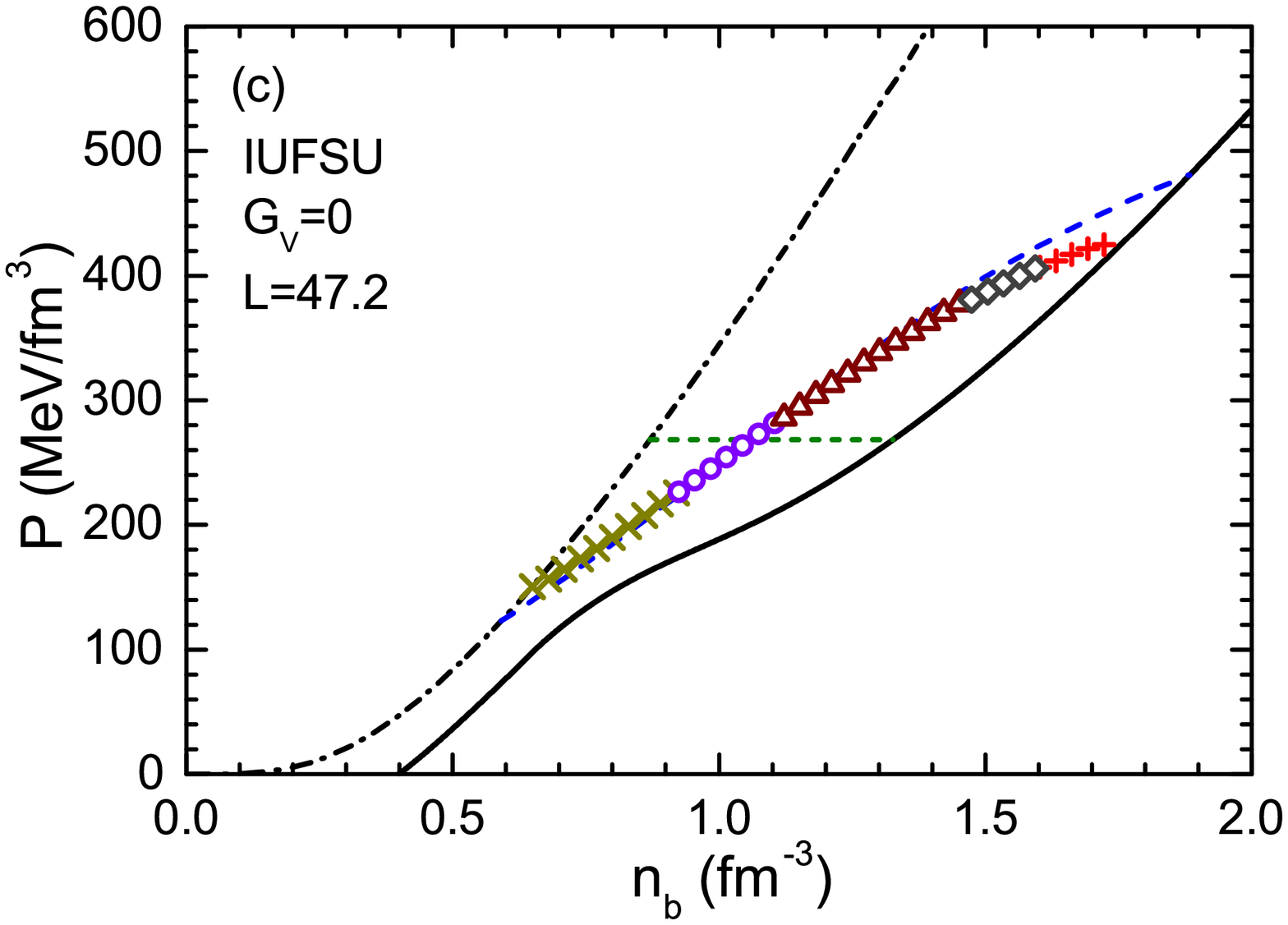}
\includegraphics[bb=30 5 580 420, width=7 cm,clip]{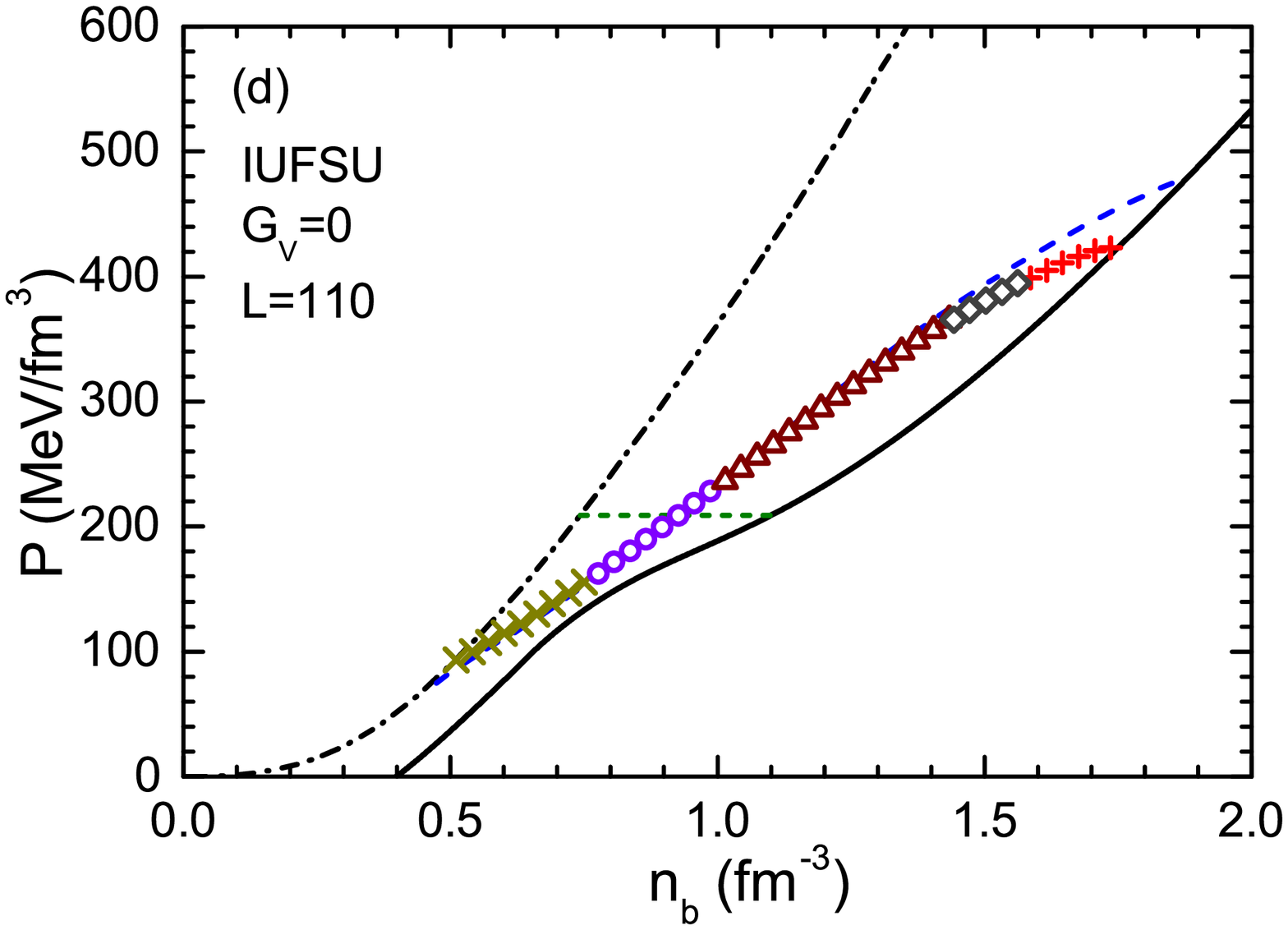}
\caption{(Color online) Pressures as a function of the baryon density
for hadronic, mixed, and quark phases.
The results of pasta phases obtained using the EM method
with $\sigma=10$ MeV/fm$^2$ are compared to those of the Gibbs and Maxwell
constructions.}
\label{fig:9nbp}
\end{figure*}

In Fig.~\ref{fig:9nbp}, we show the pressures as a function of the baryon
density for hadronic, mixed, and quark phases. The results with $L=50$ MeV and
$L=110.8$ MeV in the TM1 set are displayed in the upper panels, while those
with $L=47.2$ MeV and $L=110$ MeV in the IUFSU set are shown in the lower panels.
In the calculations, the parameters ${G_V}=0$ and $\sigma = 10$ MeV/fm$^2$
are used. The pressures of pasta phases are obtained using the EM
method, while those with the Gibbs and Maxwell constructions are shown
for comparison. It is clearly seen that the pressures of pasta phases
are very close to those of the Gibbs construction, while the pressures
of the Maxwell construction are constant shown by the green dotted lines.
The effects of symmetry energy slope $L$ on the EOS
can be observed by comparing the left and right panels. It is shown that
a smaller $L$ results in relatively larger pressures and higher onset
densities of the mixed phase. We find that qualitative behaviors of the EOS
are very similar between the TM1 and IUFSU sets, although quantitative
differences exist.

\begin{figure*}[htbp]
\includegraphics[bb=30 30 570 590, width=7 cm,clip]{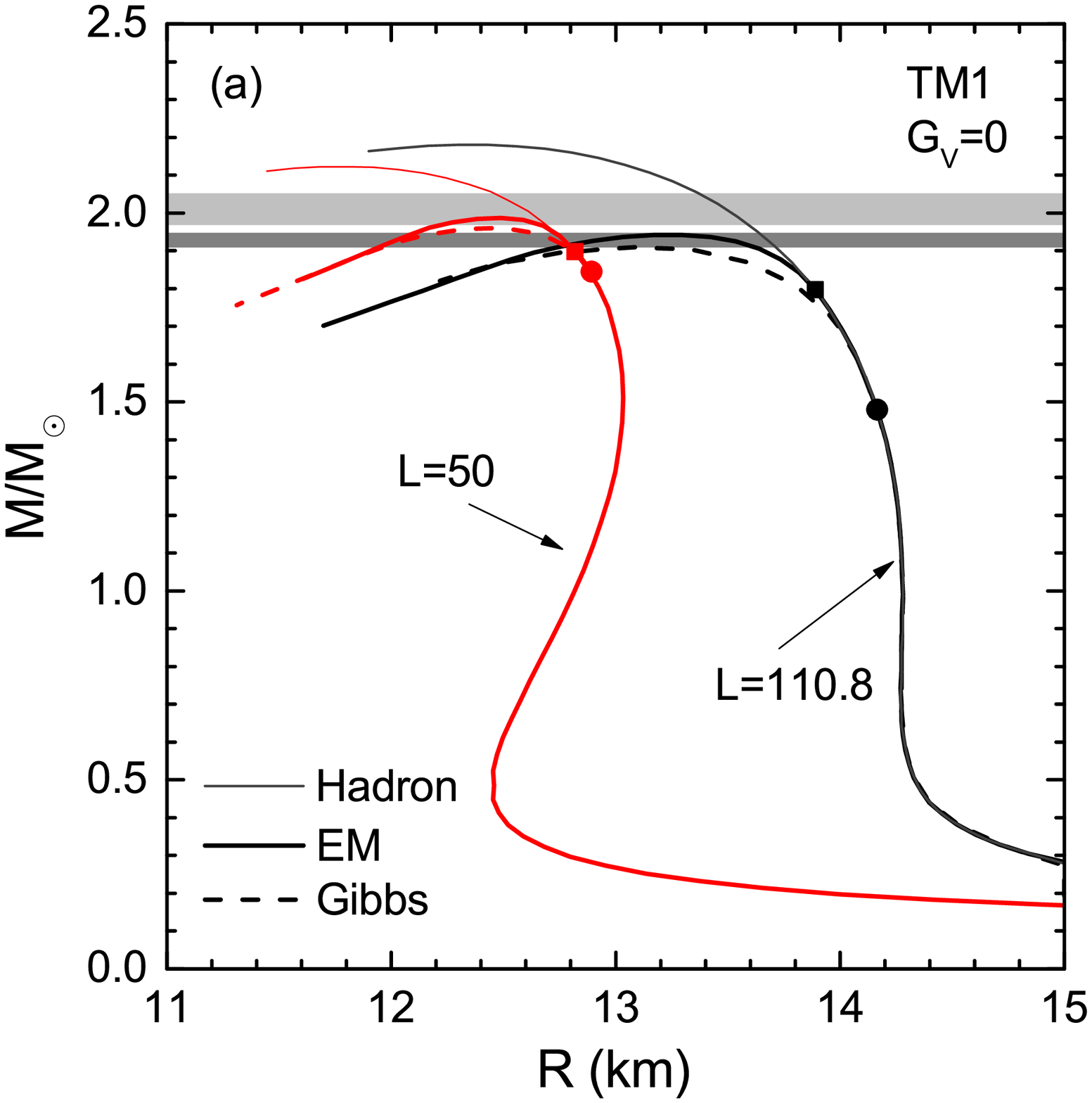}
\includegraphics[bb=30 30 570 590, width=7 cm,clip]{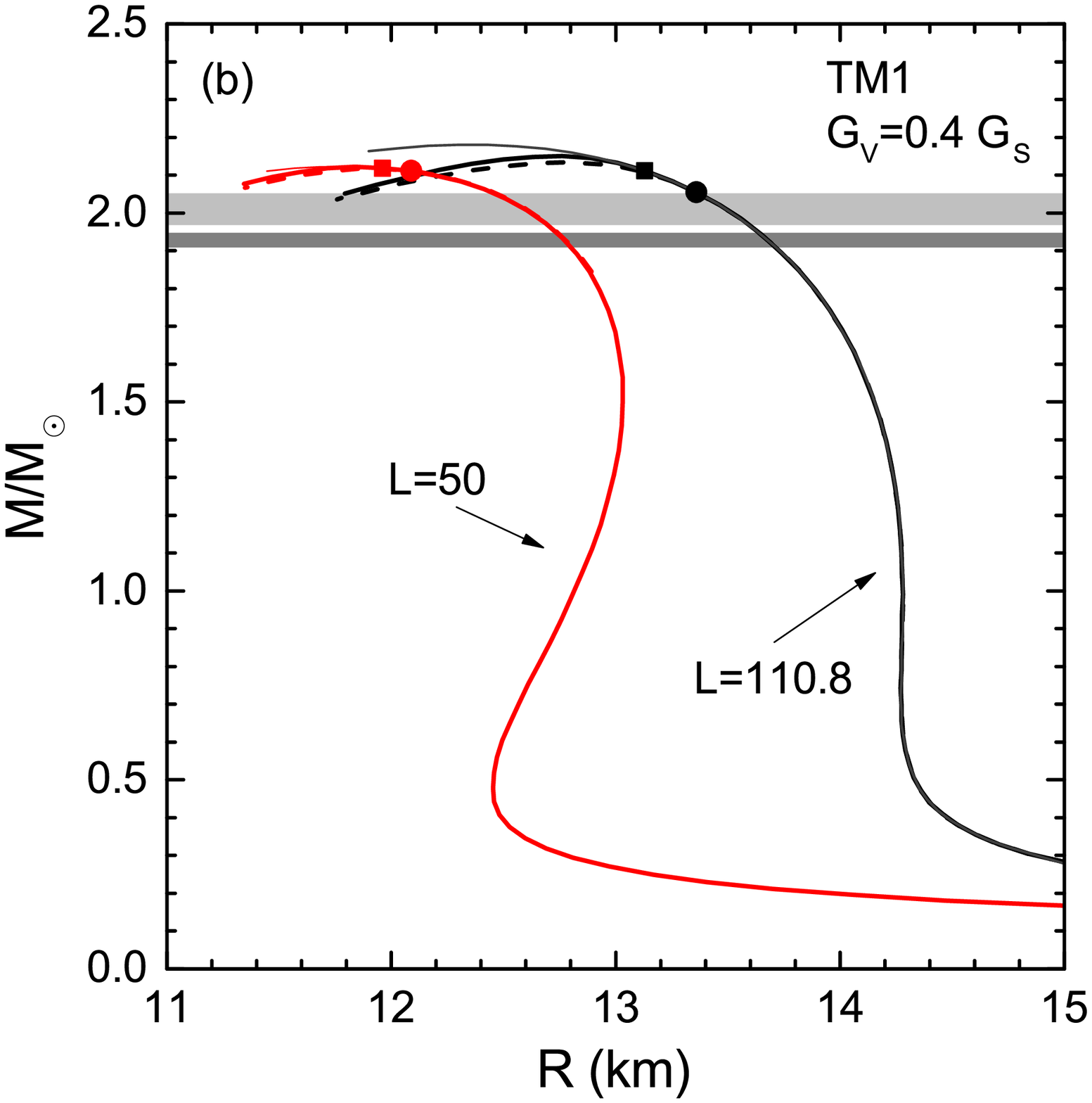}
\caption{(Color online) Mass-radius relations of neutron stars
for different EOS. The surface tension $\sigma=10$ MeV/fm$^2$ is used
in the EM method. For comparison, results obtained using pure hadronic
EOS are shown by thin solid lines. The filled squares and circles
indicate the onset of the star containing a hadron-quark mixed phase
within the EM method and Gibbs construction, respectively.
The lighter and darker shaded
regions correspond to the observational constraints of PSR J0348--0432
($M=2.01  \pm 0.04  \ M_\odot$)~\cite{Anto13} and PSR J1614--2230
($M=1.928 \pm 0.017 \ M_\odot$)~\cite{Demo10,Fons16}, respectively.}
\label{fig:10mr}
\end{figure*}

The properties of neutron stars can be obtained by solving the
Tolman-Oppenheimer-Volkoff (TOV) equation using the EOS described above,
which is matched to the low-density EOS constructed from the Thomas-Fermi
approximation within the TM1 model for the description of neutron-star crusts~\cite{Shen02}.
In Fig.~\ref{fig:10mr}, we display the mass-radius relations obtained
in the TM1 set with $L=50$ MeV and $L=110.8$ MeV,
where the observational constraints of PSR J1614--2230
($M=1.928 \pm 0.017 \ M_\odot$)~\cite{Demo10,Fons16}
and PSR J0348+0432 ($M=2.01  \pm 0.04  \ M_\odot$)~\cite{Anto13}
are shown by the darker and lighter shaded regions, respectively.
The results with ${G_V}=0$ and ${G_V}=0.4\,G_S$ are presented in the left
and right panels, respectively. For comparison, results obtained using
pure hadronic EOS are shown by thin solid lines.
It is observed that including quark degrees of freedom can soften the
EOS and reduce the maximum mass of neutron stars.
The star masses obtained using the EM method are slightly larger
than those of the Gibbs construction due to finite-size effects.
The influence of symmetry energy slope $L$ is obvious, especially on
the radius of neutron stars.
By comparing the left and right panels, we find that repulsive
vector interactions in the NJL model can significantly increase
the maximum mass of neutron stars.
To analyze neutron-star properties in more detail, we present in Table~\ref{tab:RM}
the structural properties of neutron stars with the maximum mass in several cases.
It is seen that in most cases, deconfined quarks can exist in the core of massive
stars either in mixed phase or in pure quark phase.
We emphasize that the mixed phase with the Maxwell construction is not allowed
to occur in neutron stars due to its constant pressure, but it is still
possible to form small size of pure quark matter in special cases (see Table~\ref{tab:RM})
when the surface tension is as high as required by the Maxwell construction.
On the other hand, the mixed phase with the Gibbs construction is likely present in
the interior of neutron stars, whose size depends on the vector coupling $G_V$.
The results obtained using the EM method indicate that hadron-quark pasta phases
may occur in the core of massive stars, which yield relatively larger $M_\mathrm{max}$
and smaller $R_{\mathrm{MP}}$ than those of the Gibbs construction due to
finite-size effects. It is unlikely to form pure quark matter
in neutron stars both with the Gibbs construction and in
the EM method, since the central density $n_c$ in these cases is lower than the onset of
pure quark matter. By comparing results with different values of $L$,
we can see that neutron-star structures are significantly dependent on
the symmetry energy slope $L$.
\begin{table*}[htbp]
\caption{Properties of neutron stars with the maximum mass $M_{\mathrm{max}}$.
The central baryon number density is denoted by $n_{c}$, while
$R_{\mathrm{QP}}$, $R_{\mathrm{MP}}$, and $R$ correspond to radii of
pure quark phase, mixed phase, and whole star, respectively.
The surface tension $\sigma = 10$ MeV/fm$^2$ is used in the EM method.}
\label{tab:RM}
\begin{center}
\setlength{\tabcolsep}{5.7mm}{
\begin{tabular}{llccccccc}
\hline\hline
 Model & & Method & $M_\mathrm{max}$ & $n_c$
 & $R_\mathrm{QP}$ & $R_\mathrm{MP}$ & $R$  \vspace{-0.15cm}\\
 & $L$ (MeV)  &   &$(M_\odot)$ & $(\rm{fm}^{-3})$  &(km) & (km) & (km) \\
\hline
TM1             & $L=50$       & Gibbs        & 1.96 & 0.80 & -    & 5.10 & 12.41 \\
$G_V=0$         &              & EM           & 1.98 & 0.80 & -    & 4.08 & 12.44 \\
                &              & Maxwell      & 2.04 & 0.65 & 0.25 & -    & 12.46 \\
                & $L=110.8$    & Gibbs        & 1.91 & 0.76 & -    & 7.80 & 13.09 \\
                &              & EM           & 1.94 & 0.70 & -    & 5.60 & 13.30 \\
                &              & Maxwell      & 2.04 & 0.77 & 0.82 & -    & 13.40 \\
\hline
TM1             & $L=50$       & Gibbs        & 2.12 & 0.87 & -    & 2.31 & 11.97 \\
$G_V=0.4\,G_S$  &              & EM           & 2.12 & 0.92 & -    & 1.71 & 11.84 \\
                & $L=110.8$    & Gibbs        & 2.13 & 0.80 & -    & 4.50 & 12.77 \\
                &              & EM           & 2.15 & 0.79 & -    & 3.41 & 12.77 \\
\hline
IUFSU           & $L=47.2$     & Gibbs        & 1.84 & 0.91 & -    & 4.71 & 11.64 \\
$G_V=0$         &              & EM           & 1.86 & 0.91 & -    & 3.91 & 11.67 \\
                & $L=110$      & Gibbs        & 1.80 & 0.89 & -    & 6.00 & 12.30 \\
                &              & EM           & 1.83 & 0.82 & -    & 5.11 & 12.48 \\
\hline\hline
\end{tabular}}
\end{center}
\end{table*}

\section{Conclusions}
\label{sec:6}

In this work, we studied the properties of hadron-quark mixed phase,
which may occur in the interior of massive neutron stars.
The RMF model was used to describe the hadronic phase, while the NJL model
was adopted for the quark phase.
We employed the Wigner-Seitz approximation to describe the hadron-quark
mixed phase, where coexisting hadronic and quark phases are separated
by a sharp interface.
We performed the calculations for pasta phases within the EM method,
where the equilibrium state at a given baryon density could be determined by
minimization of the total energy including surface and Coulomb contributions.
The equilibrium conditions derived in the EM method are somewhat different
from the Gibbs conditions due to finite-size effects.
A simple CP method was also used and compared for the description of
hadron-quark pasta phases, where two coexisting phases satisfy Gibbs conditions for phase
equilibrium, while the surface and Coulomb energies are perturbatively included
after the equilibrium state is achieved.
It was found that pasta structures depend on the surface tension $\sigma$,
and a smaller value of $\sigma$ could lead to more pasta configurations
during the hadron-quark phase transition.
Comparing with the results by the EM method, fewer pasta configurations
would be present and the transition density between different
shapes is independent of $\sigma$ in the CP method.
We have compared the properties of hadron-quark mixed phase obtained
from the EM and CP methods with those from the Gibbs and Maxwell constructions,
which include only bulk contributions without finite-size effects.
Since the Gibbs and Maxwell constructions correspond respectively to the two
limits of zero and very large surface tension, results of the EM and CP methods
with finite $\sigma$ were found to lie between those of the Gibbs and Maxwell
constructions.

To investigate the effects of nuclear symmetry energy on the hadron-quark phase
transition, we employed two successful RMF models (TM1 and IUFSU), which could
provide good descriptions of finite nuclei and acceptable maximum mass of neutron
stars. It was found that the IUFSU model predicted higher onset and wider range
of the mixed phase compared with the TM1 model. The qualitative
behaviors are similar between these two models. In order to examine the influence
of symmetry energy slope $L$, we adopted two sets of generated models based on
the TM1 and IUFSU parametrizations. All models in each set have the same isoscalar
saturation properties and fixed symmetry energy at the density $n_b=0.11$ $\rm{fm}^{-3}$
but have different symmetry energy slope $L$.
It has been shown that as $L$ increases, the transition densities between different
pasta configurations decrease slightly, and this tendency becomes weaker at
the end of the mixed phase. This means that the starting density of the mixed phase
is more sensitive to the value of $L$, compared with the ending density.
The influences of repulsive vector interactions in the NJL model have been
evaluated by comparing results of $G_V=0$ and $G_V=0.4\,G_S$.
The inclusion of repulsive vector interactions could significantly increase the quark
matter energy density and stiffen the EOS of quark matter. This trend becomes more apparent
with increasing density. As a result, the mixed phase with $G_V=0.4\,G_S$
would be shifted toward higher densities with a wider range, compared to the
case of $G_V=0$. Meanwhile, the critical densities of various pasta phases are
also dependent on the vector coupling $G_V$.

We calculated properties of neutron stars by using the EOS with quark degrees of
freedom. The inclusion of hadron-quark mixed phase could considerably soften the EOS
and reduce the maximum mass of neutron stars.
The star masses obtained using the EM method are slightly larger
than those of the Gibbs construction due to finite-size effects,
but lower than those of the Maxwell construction and pure hadronic matter.
The neutron-star radius is closely related to the symmetry energy slope $L$.
The repulsive vector interactions in the NJL model could significantly increase
the maximum mass of neutron stars.
Generally, there is a critical surface tension above which
the energy density of the mixed phase in the EM method is higher
than the one in the Maxwell construction, and as a result, no mixed
phase would occur inside neutron stars because the energetically favored
Maxwell construction corresponds to constant pressure.
It was found that the critical surface tension obtained using the TM1 model
is about $75$ MeV/fm$^2$ for $G_V=0$ adopted in the NJL model, while it
increases to $\approx$ 200 MeV/fm$^2$ for $G_V=0.4\,G_S$.
When a small surface tension like $\sigma = 10$ MeV/fm$^2$ was used,
we found that in most cases, hadron-quark pasta phases could occur in the core
of massive stars, but it is unlikely to form pure quark matter.
The resulting maximum masses of neutron stars are almost compatible with the
observations of PSR J1614--2230 and PSR J0348+0432.
We emphasize that both nuclear symmetry energy and repulsive vector interactions
in the NJL model can affect structural properties of neutron stars.

\section*{Acknowledgment}

This work was supported in part by the National Natural
Science Foundation of China (Grants No. 11375089 and No. 11675083).


\end{document}